\documentclass[twocolumn,pra,amsmath,amssymb,superscriptaddress]{revtex4-1}
\usepackage{graphicx,bm,dsfont,amsmath,amssymb}
\usepackage{soul}
\usepackage{color}
\usepackage{amsthm}
\usepackage{mathrsfs}
\usepackage{subfigure}
\usepackage{lipsum}
\usepackage[colorlinks=true,citecolor=blue,urlcolor=blue]{hyperref}

\newcommand{\blue}[1]{\textcolor{blue}{#1}}

\begin{document}
	
\title{Simulation of Kitaev model using one-dimensional chain of superconducting qubits and environmental effect on topological states}
\author{Yang Zhang}
\thanks{\blue{These authors contribute equally to this work}}
\affiliation{School of integrated circuits, Tsinghua University, Beijing 100084, China}
\affiliation{Frontier Science Center for Quantum Information, Beijing, China}
\author{Yun-Qiu Ge}
\thanks{\blue{These authors contribute equally to this work}}
\affiliation{School of integrated circuits, Tsinghua University, Beijing 100084, China}
\affiliation{Frontier Science Center for Quantum Information, Beijing, China}
\author{Yu-xi Liu}\email{yuxiliu@mail.tsinghua.edu.cn}
\affiliation{School of integrated circuits, Tsinghua University, Beijing 100084, China}
\affiliation{Frontier Science Center for Quantum Information, Beijing, China}

\begin{abstract}
Kitaev fermionic chain is one of the important physical models for studying topological physics and quantum computing. We here propose an approach to simulate the one-dimensional Kitaev model by a chain of superconducting qubit circuits. Furthermore, we study the environmental effect on topological quantum states of the Kitaev model. Besides the independent environment surrounding each qubit, we also consider the common environment shared by two nearest neighboring qubits. Such common environment can result in an effective non-Hermitian dissipative coupling between two qubits. Through theoretical analysis and numerical calculations, we show that the common environment can significantly change properties of topological states in contrast to the independent environment. In addition, we also find that dissipative couplings at the edges of the chain can be used to more easily tune the topological properties of the system than those at other positions. Our study may open a new way to explore topological quantum phase transition and various environmental effects on topological physics using superconducting qubit circuits.
\end{abstract}

\maketitle

\section{Introduction}

Topological quantum states are believed to be immune from small local imperfections and noises, and are thought as good candidates for encoding and manipulating quantum information to achieve fault-tolerant quantum computing~\cite{RevModPhys.82.3045,RevModPhys.83.1057,Bernevig2013robustness,Kitaev2003topological,RevModPhys.80.1083,QuantumInf.1.15001}. There has been a growing interest over recent years for simulating topological quantum properties using single superconducting quantum circuits. For example, topological invariants and topological phase transitions were experimentally simulated by single superconducting qubit circuits~\cite{Roushan2014,F2016,Ramasesh2017,Tan2019,Tao2020} via Berry phase ~\cite{Leek2007,Berger2012,Schroer2014,Zhang2017}. The space-time inversion symmetric topological semimetal~\cite{Tan2017} and topological Maxwell metal bands~\cite{Tan2017b} were also experimentally demonstrated in a single three-level superconducting quantum circuits.

Topological physics in superconducting quantum circuits with many qubits is also explored. For example, topologically protected qubits and quantum coherence were proposed by using Josephson junction arrays~\cite{IoffeNature} and superconducting qubits~\cite{NWPRL2020}. A two-component fermion model with anyonic excitations was constructed by two capacitively coupled chains of superconducting quantum circuits~\cite{Xue2009}. Quantum emulation of a spin system with topologically protected ground states was also studied~\cite{You2010}. Anyonic fractional statistical behavior is emulated in supercondcuting circuits with four qubits coupled via quantized microwave fields~\cite{Zhong2016}. Also, photon transport was observed on a Bose-Hubbard Chain~\cite{Fedorov2021} consisting of five superconducting qubits and a unit cell formed by three superconducting qubits~\cite{Roushan2016}, in which the synthetic topological materials are made of photons. Vortex-Meissner phase transition  is studied in the system of superconducting qubits~\cite{zhao2020PRA} by artificially engineering gauge potential via two-tone driving. Su-Schrieffer-Heeger (SSH)~\cite{Su1979} models were simulated by a superconducting qubit chain~\cite{Guxiu2017}, generalized SSH~\cite{Mei2018} models were also studied in superconducting qubit circuits, the edge states in the SSH chain with five qubits were experimentally observed~\cite{LuyanSun}.

It is well known that Majorana zero-energy modes play a very important role in studying topological physics and topological quantum computing. These modes are ideally supposed to be localized at edges of the system as edge states in the SSH model~\cite{Su1979}. Researches on finding Majorana modes and Majorana fermions have attracted extensive attentions in the systems of condensed matter~\cite{PhysRevLett.100.096407,PhysRevLett.105.077001,PhysRevLett.105.177002,Nature.464.187,Science.336.1003,NanoLett.12.6414,NatPhys.13.563,NatMat.14.400,PhysRevB.82.094504,PhysRevB.86.220506,PhysRevB.97.155425,PhysRevB.87.094518,PhysRevLett.122.147701,PhysRevB.87.024515,Nature.531.206,PhysRevLett.118.137701,Science.357.6348,PhysRevLett.118.137701,NatPhys.6.336,NatRevMater.3.52}. A basic model that possesses Majorana modes is Kitaev fermion chains~\cite{PhysUsp44.131}. In such model, the topology is characterized by the existence of twofold degenerate zero-energy modes, called as Majorana modes~\cite{Majorana1937}. These modes are topologically protected and believed to be robust against the imperfections and noises~\cite{PhysRevB.86.205135,PhysRevB.85.035110,NewJPhys.13.065028,PhysRevB.90.014507}. The Kitaev chains have been emulated by several physical systems, e.g., trapped ions~\cite{NewJPhys.13.115011}, optomechanical~\cite{OptExpress.26.16250} and photonic systems~\cite{GuoNC}. The digital emulation on Majorana modes was experimentally demonstrated in superconducting qubit systems~\cite{Huang2021}. However, the construction of the model Hamiltonian for the Kitaev fermion chain~\cite{PhysUsp44.131} is not realized by superconducting qubit circuits. The main challenge is how to obtain a Hamiltonian which includes the terms representing the superconducting gap. Such terms correspond to the counter-rotating part in the spin-spin coupling and is neglected via the rotating wave approximation. By using the flexible design, we here solve the problem of the counter-rotating term and propose an approach to construct Kitaev chains via superconducting quantum circuits~\cite{Nature.453.1031,Younature,Science.339.1169,PhysRep.718.1}.

Moreover, the environmental effects on topological states and phases have recently been revisited and received increasing interest~\cite{PhysRevA.94.022119,PhysRevA.95.053626,SciRep.10.6807,PhysRevLett.115.040402,PhysRevA.92.012116}. Many efforts have been made to study the dissipative effects which originate from the uncorrelated local environment~\cite{Hamazaki2019manybody,PhysRevA.98.013628,PhysRevLett.102.065703,PhysRevLett.115.040402,Xiao2017,Leykam2021}. However, it is worth noting that the common environment exists extensively in the many-body systems~\cite{PhysRevLett.127.250402,PhysRevApplied.15.044041}. Such environmental effect is equivalent to an effective Hamiltonian description which consists of both the non-Hermitian local dissipative potential and the non-Hermitian dissipative coupling between sites~\cite{Nature594.369}. Different from conventional coherent coupling, the non-Hermitian dissipative coupling possesses different physical properties, and has been applied to study level attraction~\cite{PRL121Harder,PRA13Zhao,PRA102Peng,PRA104Jiang}, light amplification and absorption~\cite{PhysRevX.5.021025,PhysRevLett.122.143901,Wanjura2020}. However, the effect of common environment on topological states in the Kitaev model is not studied, thus we will also study how common environment affects the topological states after the Kitaev model is constructed by superconducting qubit circuits.

This paper is organized as follows: In Sec.~\ref{sec2}, we first design a controllable superconducting quantum circuits which are described via spin-spin interaction, then we map the spin model to Kitaev fermion chain via the Jordan-Wigner transformation. In Sec.~\ref{sec3}, we derive an effective Hamiltonian when the common environment is included, the effective Hamiltonian resulted from the common environment is equivalent to adding a non-Hermitian dissipative coupling term to the conventional Kitaev model. In Sec.~\ref{sec4}, topological properties of the system are studied. We first study the Kitaev model when environmental effect is neglected in Sec.~\ref{sec4a}, then we consider the effect of dissipative coupling and study the following three non-Hermitian models resulted from three different dissipative mechanisms: (i) dissipative couplings exist in all neighboring qubits when the phase factor $\theta=0$ of the hopping parameter in Sec.~\ref{subsub4b1} or $\theta\neq0$ in Sec.~\ref{subsub4b2}; (ii) dissipative couplings exist in partial neighboring qubits when $\theta=0$ in Sec.~\ref{subsub4b3}. The effect of onsite dissipative potential induced by independent and common environments and comparison between dissipative coupling and onsite dissipative potential are discussed in Sec.~\ref{subsub4b4}. Finally, in Sec.~\ref{sec5}, we summarize the main results and provide further discussions on experimental feasibility in the superconducting quantum circuits.

\section{Kitaev model based on superconducting quantum circuits \label{sec2}}

\begin{figure}
\includegraphics[scale=0.38]{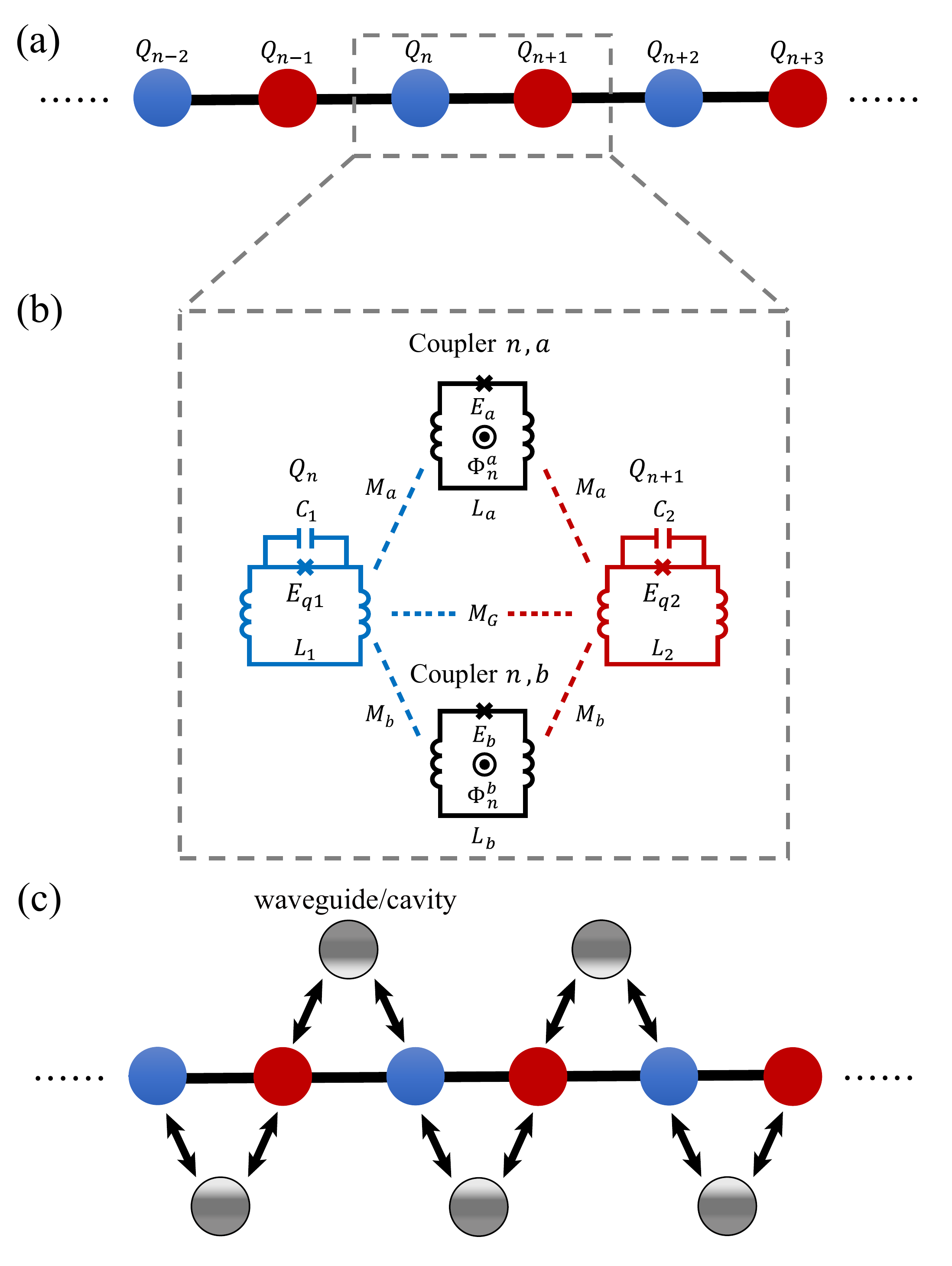}
\caption{(a) Schematic diagram of the topological qubit chain, which is made up of alternative arrangements of two kinds of qubits represented by blue and red balls. (b) Implementation of the basic structure of the above qubit chain by the superconducting qubit circuits. (c) Schematic diagram of the qubit chain with dissipative couplings, the nearest neighbor qubits are coupled to their common environments which can be mimicked by either microwave cavities or waveguides, the common environments are represented by gray balls.}\label{Fig1}
\end{figure}

As schematically shown in Fig.~\ref{Fig1}(a), a one-dimensional topological chain consists of the couplers and superconducting qubits, represented by black line segments and colored balls with the labels $Q_{n}$ ($n=1,\,2,\cdots, N$), respectively. We assume that the qubits represented by the same color balls have the same parameters. That is, all qubits represented by blue balls have the capacitance $C_{1}$, Josephson energy $E_{q1}$ and linear inductance $L_{1}$ for the odd number $n$ in the subscripts of $Q_{n}$. For qubits represented by red balls, they have the capacitance $C_{2}$, Josephson energy $E_{q2}$ and linear inductance $L_{2}$ for the even number $n$ in the subscripts of $Q_{n}$. Here, for concreteness of the following study, the chain is assumed to be formed by phase qubits coupled through rf SQUID couplers~\cite{PhysRevLett.104.177004,PhysRevLett.112.123601,PhysRevLett.98.177001}. We note that the single rf SQUID coupler between two phase qubits in Refs.~\cite{PhysRevLett.104.177004,PhysRevLett.112.123601,PhysRevLett.98.177001}  has been replaced by two rf SQUIDs for achieving our goal. It should be pointed that the chain can also be formed by other superconducting qubits.

Actually, the nearest neighboring qubits, e.g., $Q_{n}$ and $Q_{n+1}$ in Fig.~\ref{Fig1}(b), have two kinds of couplings: (i) the indirect couplings induced by $a$ and $b$ types of rf SQUID couplers $n, a$ and $n, b$; (ii) the direct coupling through their geometric mutual inductance. For simplicity, the rf SQUID couplers with the index $a$ ($b$) are assumed to be identical and have the Josephson energy $E_{a}$ ($E_{b}$) and linear inductance $L_{a}$ ($L_{b}$). In principle, the geometric mutual inductance exists between any two qubits in the qubit chain. However, considering that mutual inductance diminishes rapidly with geometric distance between qubits, here we only take the mutual inductance $M_{G}$ between the nearest neighbor qubits into account.

Applying the canonical quantization method, the magnetic flux, i.e., $\{\hat{\Phi}_{n}\}$, is considered as the canonical coordinate, and the corresponding canonical momentum is denoted by $\{\hat{P}_{n}\}$, they satisfy the canonical commutation relation $[\hat{P}_{n},\hat{\Phi}_{n}]=-i\hbar$. The total Hamiltonian of the qubit chain reads
\begin{equation}
	\begin{split}
		\hat{H}_{\rm tot}=&\sum_{n=1}^{N}\!\Big[ \frac{\hat{P}_{n}^{2}}{2C_{n}}\!+\!\frac{\hat{\Phi}_{n}^{2}}{2L_{n}}\!-\!E_{qn}\cos\left(\dfrac{2\pi}{\Phi_{0}}\hat{\Phi}_{n}\right)\!\\
		&+\! M_{n}^{\rm tot}\frac{\hat{\Phi}_{n}\hat{\Phi}_{n+1}}{L_{1}L_{2}}\Big],
	\end{split}
	\label{Eq1}
\end{equation}
in which $\Phi_{0}=h/2e$ is the magnetic flux quantum. We emphasize that $C_{n}=C_{1},\, L_{n}=L_{1}$ and $E_{qn}=E_{q1}$ when $n$ is an odd number. While $C_{n}=C_{2},\, L_{n}=L_{2}$  and $E_{qn}=E_{q2}$ when $n$ is an even number. The total mutual inductance consists of two parts $M_{n}^{\rm tot}=M_{G}+M_{n}^{\rm eff}$, here $M_{n}^{\rm eff}$ represents the effective mutual inductance between qubits $Q_{n}$ and $Q_{n+1}$ mediated by couplers, which can be continuously regulated in a wide range~\cite{PhysRevLett.104.177004,PhysRevLett.112.123601,PhysRevLett.98.177001}. Taking couplers shown in Fig.~\ref{Fig1}(b) as an example, we obtain the total flux bias applied to coupler $n,\xi$ as
\begin{equation}
	\Phi_{n}^\xi=\Phi_{0}^\xi+\delta\Phi^{\xi}\cos\left(\Omega_{\xi}\tau+\theta_{n}^\xi\right),\quad \xi=a,b,
	\label{Eq2}
\end{equation}
where $\Phi_{0}^\xi$ and $\delta\Phi^{\xi}\cos(\Omega_{\xi}\tau+\theta_{n}^\xi)$ are the $\rm dc$ and $\rm ac$ magnetic fluxes, respectively. $\Omega_{\xi}$ and $\theta_{n}^\xi$ are the frequency and initial phase of the $\rm ac$ magnetic flux. Then, the current in coupler $n,\xi$ can be represented as
\begin{equation}
	I_{n}^\xi=I_{0}^{\xi}\sin\left[\dfrac{2\pi}{\Phi_{0}}\left(\Phi_{n}^\xi-L_{\xi} I_{n}^\xi\right)\right],
	\label{Eq3}
\end{equation}
where $I_{0}^{\xi}$ is the critical current of the Josephson junction in coupler $n,\xi$. Therefore, we can get the effective mutual inductance mediated by couplers as follows:
\begin{equation}
	\begin{split}
		M_{n}^{\rm eff}=&\sum_{\xi=a,b}M_{\xi}^2\frac{\partial I_{n}^{\xi}}{\partial\Phi_{n}^{\xi}}
		\\=&\sum_{\xi=a,b}\frac{M_{\xi}^2}{L_{\xi}}\frac{\alpha_{\xi}\cos[2\pi(\Phi_{n}^\xi-L_{\xi} I_{n}^\xi)/\Phi_{0}]}{1+\alpha_{\xi}\cos[2\pi(\Phi_{n}^\xi-L_{\xi} I_{n}^\xi)/\Phi_{0}]},\\ \alpha_{\xi}=&\ \dfrac{2\pi I_{0}^{\xi}L_{\xi}}{\Phi_{0}}.
	\end{split}
	\label{Eq4}
\end{equation}
Here, we have assumed that the mutual inductance between neighboring qubits and the coupler $n,\xi$ is the same and denoted as $M_{\xi}$.

Under the weak ac flux bias condition, i.e., $\delta\Phi^{\xi}\ll\Phi_{0}^\xi$, Eq.~(\ref{Eq4}) can be expressed by a Taylor-series centered at $\Phi_{0}^\xi$ which can be written as
\begin{equation}
	M_{n}^{\rm eff}=\sum_{r=0}^{\infty}\frac{1}{r!}\dfrac{\partial^{r} M_{n}^{\rm eff}}{\partial (\Phi_{n}^{\xi})^{r}}\Big|_{\Phi_{n}^{\xi}=\Phi_{0}^{\xi}}(\Phi_{n}^{\xi}-\Phi_{0}^{\xi})^{r}.
	\label{Eq5}
\end{equation}
Truncating to the first order term of the parameter $\delta\Phi^\xi$, we have
\begin{equation}
	\begin{split}
		M_{n}^{\rm eff}&\approx\sum_{\xi=a,b}\Big[M_{\xi}^{0}+\delta M_{\xi}\cos(\Omega_{\xi}\tau+\theta_{n}^\xi)\Big],\\
		M_{\xi}^{0}&=\frac{M_{\xi}^2}{L_{\xi}}\frac{\alpha_{\xi}\cos(\beta_{\xi})}{1+\alpha_{\xi}\cos(\beta_{\xi})},\\
		\delta M_{\xi}&=-\frac{M_{\xi}^2}{L_{\xi}}\frac{2\pi\alpha_{\xi}\sin(\beta_{\xi})}{[1+\alpha_{\xi}\cos(\beta_{\xi})]^{3}}\frac{\delta\Phi^\xi}{\Phi_{0}},\\
		\beta_{\xi}&=\dfrac{2\pi(\Phi_{0}^{\xi}-L_{\xi}I_{n}^{\xi}|_{\Phi_{n}^{\xi}=\Phi_{0}^{\xi}})}{\Phi_{0}}.
	\end{split}
	\label{Eq6}
\end{equation}
Furthermore, to avoid the effect of the geometric mutual inductance $M_G$ on topological properties of the system, we assume
\begin{equation}
	M_{G}+\sum_{\xi=a,b} M_{\xi}^{0}= 0.
	\label{Eq7}
\end{equation}
Based on recent experimental results~\cite{PhysRevB.80.052506,PhysRevApplied.13.034037}, all the above conditions can be well satisfied.

We now project the Hamiltonian in Eq.~(\ref{Eq1}) to the qubit basis. Here we can define the ladder operators, i.e., $\hat{\sigma}_{n}^{\dag}=|1\rangle\!\ \!\!_{nn}\!\langle0|,\  \hat{\sigma}_{n}^{-}=|0\rangle\!\ \!\!_{nn}\!\langle1| $  for $n$th qubit $Q_{n}$ with its ground $|0\rangle_{n}$ and the first excited state $|1\rangle_{n}$. Then the Hamiltonian of the qubit chain in Eq.~(\ref{Eq1}) can be rewritten as
\begin{equation}
	\begin{split}
		\hat{H}_{\rm tot}\!=\!&\:\,\underset{n=1}{\overset{N}{{\displaystyle\sum}}}\Big[\hbar\Omega_{n}\hat{\sigma}_{n}^{\dag}\hat{\sigma}_{n}^-\!\\
		&+\sum_{\xi=a,b}\!\hbar J_{\xi}\cos(\Omega_{\xi}\tau\!+\!\theta_{n}^{\xi})(\hat{\sigma}_{n}^{-}\hat{\sigma}_{n+1}^{-}\!+\!\hat{\sigma}_{n}^{-}\hat{\sigma}_{n+1}^{\dag}\!+\!{\rm H.c.})\Big]
	\end{split}
	\label{Eq8}
\end{equation}
in the qubit basis. Here, we note that $\Omega_{n}\equiv\Omega_{1}$ when $n$ is an odd number, and $\Omega_{n}\equiv\Omega_{2}$ when $n$ is an even number. The analytical forms of the parameters $\Omega_{n}$ and $J_{\xi}$ are approximately given below.

The analytical form of the eigenstates of the qubit, e.g., $n$th qubit $Q_n$, cannot be obtained exactly. However, we can derive an approximate solution by using the perturbation theory. That is, we expand the cosine function $\cos(\hat{\phi}_n)$ to the fourth order of $\hat{\phi}_n= 2\pi\hat{\Phi}_n/\Phi_{0}$, i.e., $\cos(\hat{\phi}_n)\approx1-\hat{\phi}_n^2/2+\hat{\phi}_n^4/24$, and treat the fourth-order nonlinear term as a perturbation. We define $|j\rangle_{n}^{(0)}$ as the $j$th excited state of the $n$th qubit when the high-order anharmonic terms of the qubit are neglected, the corresponding eigenenergies $E_{n,j}^{(0)}$ are
\begin{equation}
	\begin{split}
		E_{n,j}^{(0)}&=(j+\frac{1}{2})\hbar\omega_{n},\quad j=0,\,1,\,2,\cdots,\\
		\omega_{n}&=\left\{
		\begin{aligned}
			&\omega_{1},\quad \rm for\  \rm odd\ number\  n,\\
			&\omega_{2},\quad \rm for\  \rm even\ number\  n,
		\end{aligned}
		\right.\\		
		\omega_{i}=\
		&\dfrac{1}{\sqrt{l_{i}C_{i}}}, \quad  l_{i}=\dfrac{\Phi_{0}^{2}L_{i}}{\Phi_{0}^{2}+4\pi^{2}E_{qi}L_{i}}, \quad i=1,\,2.
	\end{split}
\label{Eq9}
\end{equation}
The first-order correction to $|j\rangle_{n}^{(0)}$, arising from the perturbation term $-E_{qn}\hat{\phi}_n^4/24$, is given as
\begin{equation}
	\begin{split}
		|j\rangle_{n}^{(1)}=-\frac{E_{qn}}{24}\underset{f\neq j}{{\displaystyle\sum}}\dfrac{\ \ _{n}^{(0)}\langle f|\hat{\phi}_n^4|j\rangle_{n}^{(0)}}{E_{n,j}^{(0)}-E_{n,f}^{(0)}}|f\rangle_{n}^{(0)}.
	\end{split}
\label{Eq10}
\end{equation}
Then, we can obtain the normalized eigenstates under the first-order approximation as
\begin{equation}
	\begin{split}
		|0\rangle_{n}&=\dfrac{|0\rangle_{n}^{(0)}+6\sqrt{2}g_{n}^{i}|2\rangle_{n}^{(0)}+\sqrt{6}g_{n}^{i}|4\rangle_{n}^{(0)}}{\sqrt{1+78(g_{n}^{i})^2}},\\
		|1\rangle_{n}&=\dfrac{|1\rangle_{n}^{(0)}+10\sqrt{6}g_{n}^{i}|3\rangle_{n}^{(0)}+\sqrt{30}g_{n}^{i}|5\rangle_{n}^{(0)}}{\sqrt{1+630(g_{n}^{i})^2}},\\
		g_{n}^{i}&=\left\{
		\begin{aligned}
			&g_{1},\quad \rm for\  \rm odd\ number\  n,\\
			&g_{2},\quad \rm for\  \rm even\ number\  n,
		\end{aligned}
		\right.\\
		g_{i}&=\dfrac{e^{4}l_{i}E_{qi}}{12\hbar^{3}\omega_{i}C_{i}},\quad \quad \quad i=1,2,\\
	\end{split}
	\label{Eq11}
\end{equation}
in which the value of parameters $g_{i}$ are about  $g_{i}\approx10^{-3}$~\cite{PhysRevLett.121.157701,PhysRevLett.114.010501,PhysRevLett.125.267701}. For the transmon and phase qubits, Eq.~(\ref{Eq11}) agrees roughly with numerical ones~\cite{PRA76}. Moreover, under the first-order approximation, we can further obtain the analytical form of the parameters in Eq.~(\ref{Eq8}) as
\begin{equation}
	\begin{split}
		\Omega_{i}=&\ \omega_{i}(1-24g_{i}),\\
		J_{\xi}=&\prod_{i=1}^{2}\sqrt{\frac{\delta M_{\xi}\omega_{i}l_{i}(1+12g_{i}+510g_{i}^{2})^2}{2L_{i}^2(1+78g_{i}^{2})(1+630g_{i}^{2})}},
	\end{split}
	\label{Eq12}
\end{equation}
with $i = 1, 2$. If the frequencies and phases of the $\rm ac$ magnetic fluxes applied to the couplers satisfy the following conditions
\begin{equation}
	\begin{split}
		\Omega_{a}&=\Omega_{2}-\Omega_{1}, \quad \quad \quad \quad\theta_{n}^a = (-1)^n\theta,\\ \Omega_{b}&=\Omega_{1}+\Omega_{2}+2\dfrac{\mu}{\hbar},\quad\:\:\:\:\!\!\!\:\! \theta_{n}^b = 0,
	\end{split}
	\label{Eq13}
\end{equation}
with the constant $\mu$, then in the rotating reference frame $\hat{U}(\tau)=\exp(-i\hat{H}_{\rm free}\tau/\hbar)$ with
\begin{equation}
	\hat{H}_{\rm free}=\sum_{n=1}^{N}\Big[\hbar\left(\Omega_{n}+\dfrac{\mu}{\hbar}\right)\hat{\sigma}_{n}^{\dag}\hat{\sigma}_{n}\Big],
	\label{Eq14}
\end{equation}
the effective Hamiltonian can be expressed as
\begin{equation}
	\begin{split}
		\hat{H}_{\rm rot}=&\ \  \hat{U}^{\dagger}(\tau)\hat{H}_{\rm tot}\hat{U}(\tau)-i\hbar\hat{U}^{\dagger}(\tau)\dfrac{\partial\hat{U}(\tau)}{\partial \tau}\\=&\ \underset{n=1}{\overset{N}{{\displaystyle\sum}}}\Big[-\mu \hat{\sigma}_{n}^{\dag}\hat{\sigma}_{n}^{-}-t(\hat{\sigma}_{n+1}^{\dag}\hat{\sigma}_{n}^-e^{i\theta}+{\rm H.c.})   \\
		&\ \:\! \, \:\! + \Delta(\hat{\sigma}_{n}^-\hat{\sigma}_{n+1}^-+{\rm H.c.})\Big],
	\end{split}
	\label{Eq15}
\end{equation}
with
\begin{equation*}
	t=-\dfrac{\hbar J_{a}}{2},\quad\quad \Delta=\dfrac{\hbar J_{b}}{2}.
\end{equation*}
We can derive the same effective Hamiltonian as in Eq.~(\ref{Eq15}) with the gauge transformation if the initial phase $\theta_{n}^b \neq 0$. It is clear that the parameters $\mu$,  $te^{i\theta}$ and $\Delta$ can be adjusted independently by the magnetic fluxes applied to the couplers. Let us now apply the Jordan-Wigner transformation
\begin{equation}
	\hat{\sigma}_{n}^\dagger = \exp\left( -i\pi\sum_{k=1}^{n-1} \hat{a}_k^\dagger \hat{a}_k \right) \hat{a}_n^\dagger, \quad \hat{\sigma}_{n}^z = 2\hat{a}_n^\dagger \hat{a}_n - 1.
	\label{Eq16}
\end{equation}
with fermionic creation (annihilation) operators $\hat{a}_{n}^{\dagger} (\hat{a}_{n})$~\cite{Introduction.coleman} to Eq.~(\ref{Eq15}), then Eq.~(\ref{Eq15}) can be rewritten as
\begin{equation}
	\begin{split}
		\hat{H}=-&\mu\:\underset{n=1}{\overset{N}{{\displaystyle\sum}}}\,\hat{a}_{n}^{\dag}\hat{a}_{n}-t\underset{n=1}{\overset{N-1}{{\displaystyle\sum}}}(e^{i\theta}\hat{a}_{n+1}^{\dag}\hat{a}_{n}+\rm H.c.)
		\\+&\Delta\underset{n=1}{\overset{N-1}{{\displaystyle\sum}}}(\hat{a}_{n}\hat{a}_{n+1}+\rm H.c.),
	\end{split}
	\label{Eq17}
\end{equation}
which is the  Hamiltonian of the Kitaev fermion chain~\cite{PhysUsp44.131}.

\section{Environmental effects on Kitaev model} \label{sec3}

In above, we show how an equivalent Hamiltonian of the Kitaev fermion chain can be derived via an ideal chain of superconducting qubit circuits. In practise, any system inevitably interacts with its environment. In our study here, besides the local environment which acts on each qubit and results in local energy dissipation, the couplers can also induce common environment between neighboring qubits, which can usually be mimicked by the microwave cavity or waveguide in superconducting qubit systems. These common environments can result in both local energy dissipation and effective dissipation coupling between neighboring qubits~\cite{PhysRevX.5.021025}. The local energy dissipation in Kitaev model system is extensively studied, thus we here mainly focus on the nonlocal effect of the dissipative coupling induced by the common environment. For simplicity and without loss of generality, as shown in Fig.~\ref{Fig1}(c), we here assume that  the neighboring qubits share a common environment.

Therefore, when we only consider the common environment, the system dynamics is governed by the master equation
\begin{equation}
	\frac{d}{d\tau}\hat{\rho}=\dfrac{1}{i\hbar}\left[\hat{H}_{\rm rot},\hat{\rho}\right]+\mathcal{L}\hat{\rho},
	\label{Eq18}
\end{equation}
with the form of the Lindblad operator~\cite{PhysRevX.5.021025,PhysRevLett.123.127202}
\begin{equation}
	\begin{split}
		\mathcal{L}\hat{\rho}= \sum_{n=1}^{N-1}\dfrac{2\gamma}{\hbar}\mathcal{L}[ \hat{\sigma}_n^-+ \hat{\sigma}_{n+1}^-]\hat{\rho}.
		\label{Eq19}
	\end{split}
\end{equation}
The Lindblad operator $\mathcal{L}\left[\hat{o}\right]\hat{\rho}$ is defined as
\begin{equation}
	\mathcal{L}\left[\hat{o}\right]\hat{\rho}=\hat{o}\hat{\rho} \hat{o}^\dagger - \frac12 \hat{o}^\dagger \hat{o}\hat{\rho} -\frac12 \hat{\rho} \hat{o}^\dagger \hat{o}.
	\label{Eq20}
\end{equation}
The parameter $\gamma$ characterizes the correlated decay between neighboring qubits. Then the effective non-Hermitian Hamiltonian reads
\begin{equation}
\begin{split}
\hat{H}_{\rm eff} = -( \mu + 2i\gamma) \sum_{n=1}^{N} \hat{\sigma}_n^\dagger \hat{\sigma}_n^-  + \Delta \sum_{n=1}^{N-1} \left( \hat{\sigma}_n^-\hat{\sigma}_{n+1}^- + {\rm H.c.} \right) \\
- \sum_{n=1}^{N-1} \left[ \left( te^{i\theta} + i\gamma \right)\hat{\sigma}_{n+1}^\dagger \hat{\sigma}_n^- + \left( te^{-i\theta} + i\gamma \right)\hat{\sigma}_{n}^\dagger \hat{\sigma}_{n+1}^- \right],
			\label{Eq21}
\end{split}
\end{equation}
in which the terms of dissipative coupling and local dissipative potential are  denoted by $-i\gamma(\hat{\sigma}_{n+1}^\dagger \hat{\sigma}_n^- + {\rm H.c.})$ and $-2i\gamma\hat{\sigma}_{n}^\dagger\hat{\sigma}_n^-$, respectively.  We note that  the local dissipation rate $-2i\gamma$  for the term $\hat{\sigma}_{n}^\dagger\hat{\sigma}_n^-$ can be modified when the independent environment for each qubit is included. Here, to highlight the role of dissipative coupling, all local dissipation of the qubit is first neglected in the following study. Thus, the non-Hermitian Hamiltonian  under the Jordan-Wigner transformation can be written as
\begin{equation}
	\begin{split}
		\hat{H}_{\rm nonH}\! =\! & -\mu \sum_{n=1}^{N} \hat{a}_n^\dagger\hat{a}_n\! +\! \Delta \sum_{n=1}^{N-1} \left( \hat{a}_n\hat{a}_{n+1}\! +\! {\rm H.c.} \right) \\
		&-\! \sum_{n=1}^{N-1} \left[ \left( te^{i\theta}\! +\! i\gamma \right)\!\hat{a}_{n+1}^\dagger \hat{a}_n \!+\! \left( te^{-i\theta}\! +\! i\gamma \right)\!\hat{a}_{n}^\dagger \hat{a}_{n+1} \right]\!.
		\label{Eq22}
	\end{split}
\end{equation}

\section{Topological properties \label{sec4}}

Let us now study the topological states of the Hermitian Hamiltonian $\hat{H}$ and non-Hermitian Hamiltonian $\hat{H}_{\rm nonH}$ derived in Eqs.~(\ref{Eq17}) and (\ref{Eq22}), respectively. For the Hermitian Hamiltonian $\hat{H}$ in  Eq.~(\ref{Eq17}), we mainly study the influence of the phase $\theta$ in the hopping terms on the topological properties of system. For the non-Hermitian Hamiltonian $\hat{H}_{\rm nonH}$, we will study the changes of topological states due to the dissipative couplings for three different cases.

\subsection{ Kitaev model with the Hermitian Hamiltonian  \label{sec4a}}

In this section,  the topological properties of the ideal system is discussed, i.e., the environmental effect is neglected. We will mainly discuss how the initial phases for all control fields through the couplers affect the topological properties under two cases. These initial phases can be varied by adjusting the control fields through the couplers.

\subsubsection{Case \uppercase\expandafter{\romannumeral1}: $\theta=0$ \label{subsub4a1}}

If the initial phases of all control fields through the couplers are zero, i.e., $\theta^{a}_{n}=(-1)^n\theta=0$, then the Hamiltonian $\hat{H}$ in Eq.~(\ref{Eq17}) can be reduced to
\begin{equation}
	\begin{split}
		\hat{H}_{0} =& -\mu \sum_{n=1}^{N} \hat{a}_n^\dagger\hat{a}_n - t \sum_{n=1}^{N-1} \left( \hat{a}_{n+1}^\dagger \hat{a}_n + {\rm H.c.} \right) \\
		&+\Delta \sum_{n=1}^{N-1} \left( \hat{a}_n\hat{a}_{n+1} + {\rm H.c.} \right),
		\label{Eq23}
	\end{split}
\end{equation}
 which is a standard Kitaev model and has been extensively studied~\cite{PhysUsp44.131}. It has been shown that  the topological phase transition point occurs at $\mu = 2t$ when the parameters satisfy the condition $t=\Delta$. If $\mu<2t$, the system is in topologically non-trivial phase, and  two zero-energy states corresponding to Majorana bound states (MBS) localize at the two ends of the system~\cite{NatRevPhys.2.575}. For the case of $\mu>2t$, the system is in topologically trivial phase, and there is no MBS. However, as shown in the next subsection,  these results are significantly changed when the initial phase $\theta$ is nonzero, i.e., the topological properties can be controlled by the initial phase of the control fields.

\subsubsection{Case \uppercase\expandafter{\romannumeral2}: $\theta \neq 0$ \label{subsub4a2}}

\begin{figure}
	\includegraphics[scale=0.29]{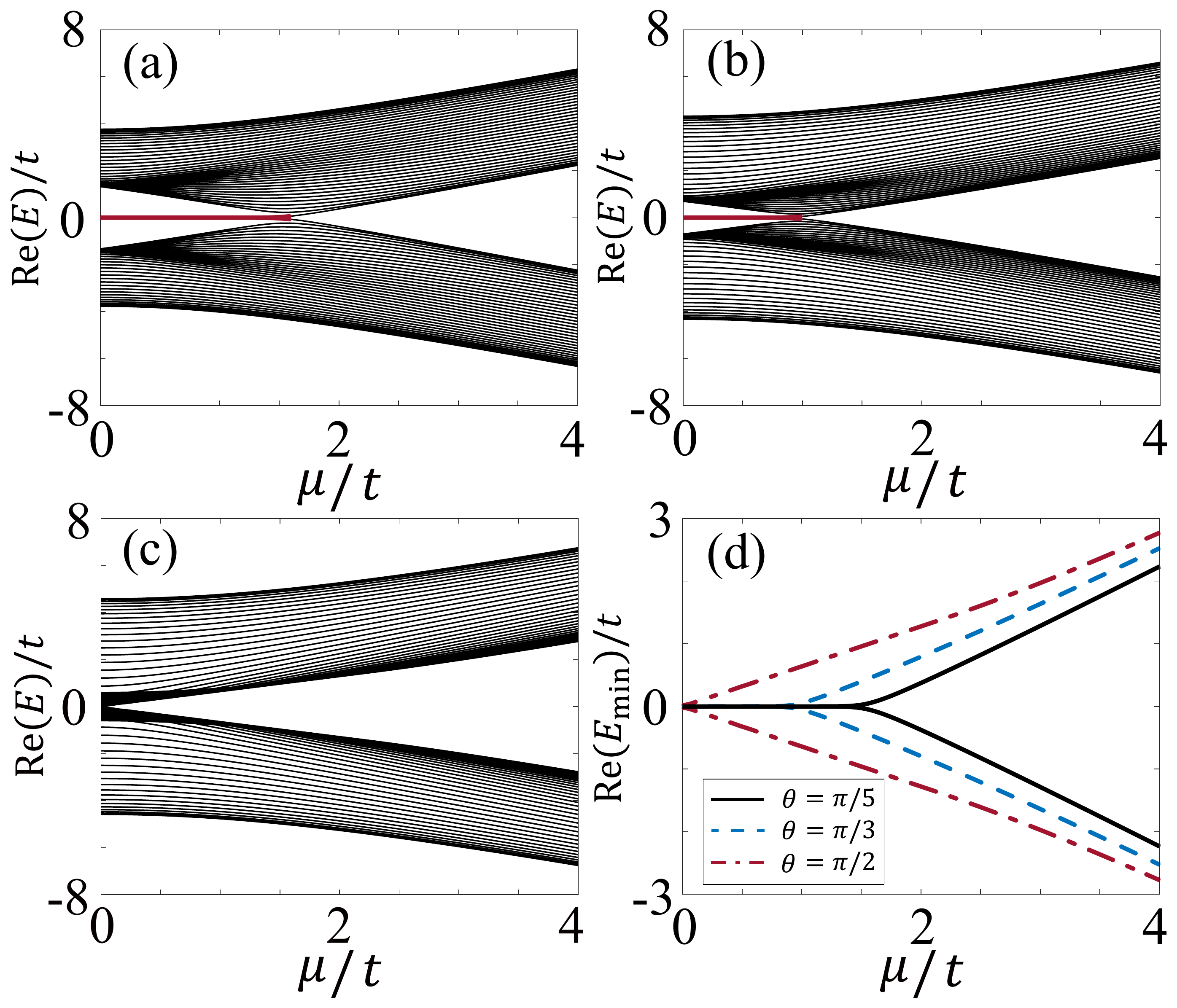}
	\caption{The real parts of energy spectrum of $\hat{H}$ as a function of $\mu$ for different $\theta$, under open boundary condition. (a) $\theta = \pi/5$, (b) $\theta = \pi/3$, and (c) $\theta = \pi/2$. Here, the red solid lines correspond to MBS. The lowest energies for these three cases are shown in (d). Other parameters chosen here are $\Delta/t=1$, $N=40$, and $t=1$.}\label{Fig2}
\end{figure}

We now study the topological properties of the system when the initial phases of  all control fields through the couplers are nonzero, i.e., $\theta^{a}_{n}=(-1)^n\theta\neq 0$. Here, we assume $t = \Delta$ in the following study. As shown in Figs.~\ref{Fig2}(a)-(d), the variations of the phase $\theta$ affect the energy spectrum and the position of the topological phase transition point. When the phase $\theta$ is changed from $0$ to $\pi/2$, we find that the phase transition point and energy spectrum gradually move toward the left of $\mu=2t$, which is the phase transition point for $\theta=0$. For example, when $\theta=\pi/5$, compared with the case of $\theta=0$~\cite{PhysUsp44.131}, the phase transition point is moved toward the left of $\mu=2t$. If $\theta$ is further increased, e.g., $\theta=\pi/3$, then the phase transition point further moves toward the left. When $\theta=\pi/2$, the phase transition point moves to the origin, that is, there is no topological phase transition and the system is in the topologically trivial phase no matter how the parameter $\mu$ varies.

 To further obtain the analytical expression of the topological phase transition point, we transform the Hamiltonian $\hat{H}$ in Eq.~(\ref{Eq17}) into the momentum space. Under periodic boundary conditions, the Hamiltonian $\hat{H}$ can be represented as
\begin{equation}
	\hat{H} =\frac{1}{2}\sum_{k} \hat{\Phi}_k^\dagger H_{k} \hat{\Phi}_k,
	\label{Eq24}
\end{equation}
with $\hat{\Phi}_k^\dagger = (\hat{a}_k^\dagger, \hat{a}_{-k})$. Here, $\hat{a}_k$ is the Fourier transformation of $\hat{a}_n$, i.e., $\hat{a}_k = \sum_n e^{-ikn}\hat{a}_n/\sqrt{N}$ and
\begin{equation}
H_{k} = h_{I}(k)I + h_{y}(k)\sigma_y + h_{z}(k)\sigma_z,
\label{Eq25}
\end{equation}
 with  the identity $I$ and Pauli matrices $\sigma_{y(z)}$ , respectively.  The partameters $h_{I}(k)$, $h_{y}(k)$ and $h_{z}(k)$ are
\begin{equation}
	\begin{split}
	h_{I}(k) &= -2t\sin k \sin \theta, \\
 	h_{y}(k) &=\:2\Delta\sin k, \\
	h_{z}(k)& = -\mu - 2t\cos k \cos \theta.
	\end{split}
\label{Eq26}
\end{equation}
Eigenvalues $E_{k}^\pm$ of $H_{k}$, which characterize the dispersion relation of the bulk, are given by
\begin{equation}
E_{k}^\pm = h_{I}(k) \pm \sqrt{h_{y}(k)^2 + h_{z}(k)^2}.
\label{Eq27}
\end{equation}
By setting $E_{k}^+=E_{k}^-=0$, we can obtain the phase transition condition when $\Delta = t$, which manifests the relation
\begin{equation}
	\mu = 2t |\cos\theta|.
	\label{Eq28}
\end{equation}
Compared with the case of the standard Kitaev model in Sec.~\ref{subsub4a1}, the topological phase transition takes place with a smaller $\mu$ when $\theta \neq 0$. The system is topologically non-trivial under the condition $\mu<2t|\cos\theta|$, and topologically trivial under the condition $\mu>2t|\cos\theta|$. Therefore, the phase $\theta$ can be considered as a control parameter for the system topological phase transition. The above results agree well with Fig.~\ref{Fig2}(d). For the case $\Delta \neq t$, Eq.~(\ref{Eq28}) needs further modification, but the conclusion remains valid, the phase $\theta$ can also control the topological property of system when $\Delta \neq t$.

\subsection{Kitaev model with Non-Hermitian dissipative couplings}

In this section, we will study the effects of the non-Hermitian dissipative couplings on the topological states of Kitaev chain. Here, we will mainly focus on following three cases: (i) the Hamiltonian with the  dissipative couplings  given by Eq.~(\ref{Eq22}) with $\theta = 0$; (ii) the Hamiltonian with the  dissipative couplings given by Eq.~(\ref{Eq22}) with $\theta \neq 0$; (iii) the Hamiltonian with the  dissipative couplings only for special positions of the Kitaev chain derived from  Eq.~(\ref{Eq22}) with $\theta = 0$.

\begin{figure}
	\includegraphics[scale=0.33]{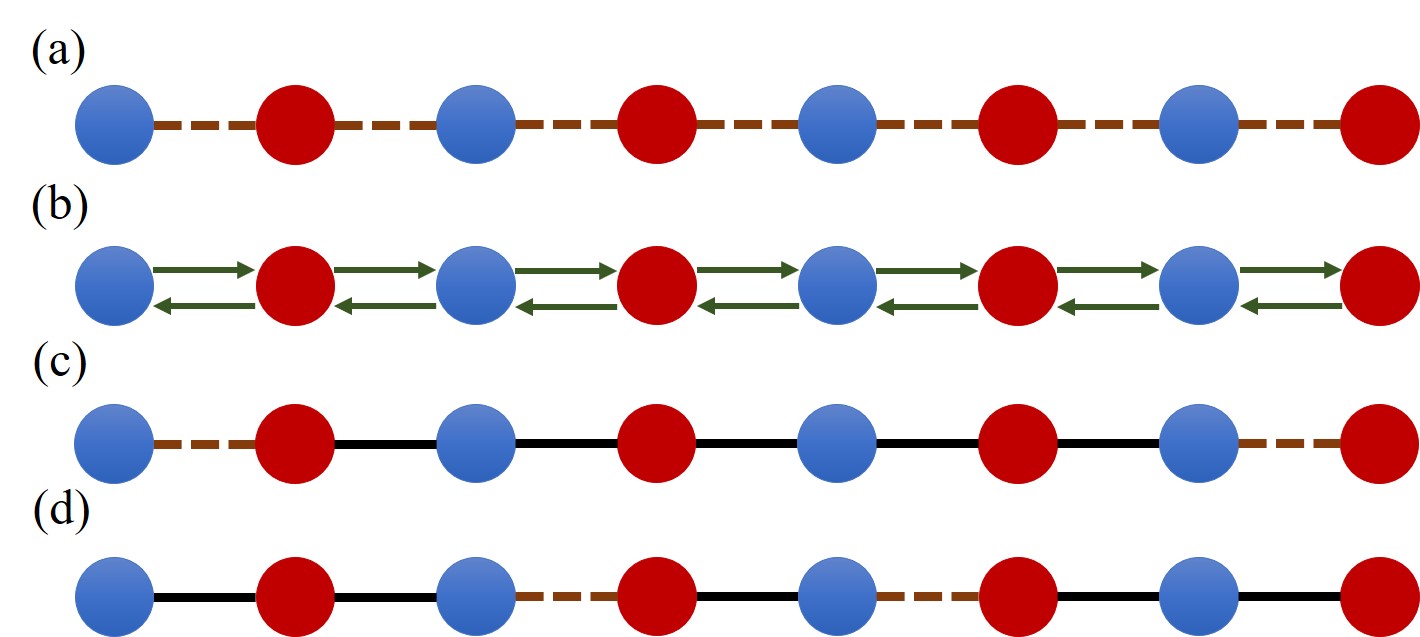}
	\caption{Schematic diagrams of several non-Hermitian Kitaev models that we mainly study, where the blue circles represent the qubits. (a) Dissipative couplings are introduced into all neighboring qubits when $\theta=0$. (b) Dissipative couplings are introduced into all neighboring qubits when $\theta\neq0$. Noting that the couplings between qubits are nonreciprocal in this case. (c) Dissipative couplings are introduced at the boundaries of qubit chain. (d) Dissipative couplings are introduced at certain positions of qubit chain. The black solid lines, dark brown dotted lines, left arrows and right arrows indicate different types of interactions with hopping parameters $t$, $t+i\gamma$, $te^{-i\theta}+i\gamma$ and $te^{i\theta}+i\gamma$, respectively.}\label{Fig3}
\end{figure}

\subsubsection{Topology of the Kitaev model with the dissipative couplings for $\theta = 0$ \label{subsub4b1}}

We first study the case that there are dissipative couplings for all nearest neighboring qubits when $\theta = 0$, which is schematically shown in Fig.~\ref{Fig3}(a). The Hamiltonian of the system can be derived from Eq.~(\ref{Eq22}) with  $\theta = 0$ and is given as
\begin{equation}
	\begin{split}
		\hat{H}_1 =&\ \, \hat{H}_{0} + \hat{U}_1,\\
	\hat{U}_1 =& -i\gamma \sum_{n=1}^{N-1} \left( \hat{a}_{n+1}^\dagger \hat{a}_n + {\rm H.c.} \right),
	\end{split}
	\label{Eq29}
\end{equation}
with $H_{0}$ given in Eq.~(\ref{Eq23}). It is clear that the Hamiltonian in Eq.~(\ref{Eq29}) is a special case of the Hamiltonian $\hat{H}_{\rm nonH}$ in Eq.~(\ref{Eq22}) and has a complex energy spectrum. Moreover, the MBS are very sensitive to the couplings between qubits, thus the dissipative coupling term $\hat{U}_{1}$ will inevitably affects MBS.

\begin{figure}
	\includegraphics[scale=0.2]{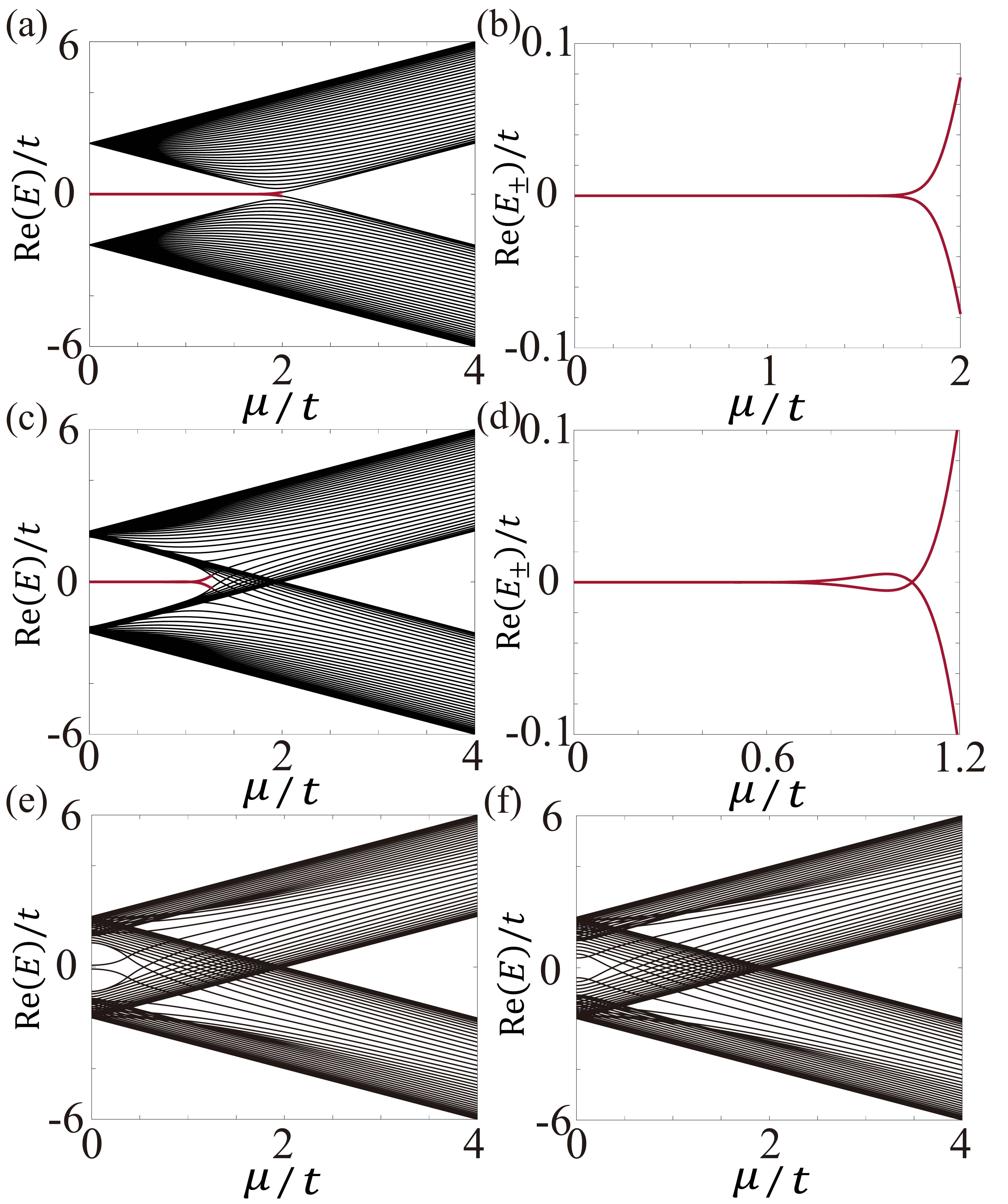}
	\caption{The real parts of energy spectrum of the non-Hermitian Kitaev model $\hat{H}_1$ as a function of $\mu$ for different $\gamma$, under open boundary condition. (a) and (b) $\gamma/t = 0$, (c) and (d) $\gamma/t=1$, (e) $\gamma/t=3$,  and (f) $\gamma/t=4$. (b) and (d) are the enlarged figures of the red solid curves in (a) and (c). Other parameters are the same as those in Fig.~\ref{Fig2}.}\label{Fig4}
\end{figure}

Under open boundary condition (OBC), the Hamiltonian $\hat{H}_1$ can be represented as $\hat{H}_1 = \frac{1}{2} \hat{\Psi}^\dagger H_{1}^{(0)} \hat{\Psi}$, in which
\begin{equation}
	\begin{split}
	\hat{\Psi} = \left( \hat{a}_1, \cdots, \hat{a}_{N}, \hat{a}_1^\dagger, \cdots, \hat{a}_N^\dagger \right)^T\!,\quad H_{1}^{(0)}=\begin{bmatrix}
			C & S \\
			S^\dagger & -C
		\end{bmatrix}.
	\end{split}
	\label{Eq30}
\end{equation}
Here, the concrete forms of matrices $C$ and $S$ are
\begin{equation}
	\begin{split}
		C=&\begin{bmatrix}
			-\mu      & -t^\prime &           &           &           &           &           \\
			-t^\prime & -\mu      & -t^\prime &           &           &           &           \\
			& -t^\prime & -\mu      & -t^\prime &           &           &           \\
			&           &   \ddots  &  \ddots   &  \ddots   &           &           \\
			&           &           & -t^\prime &   -\mu    & -t^\prime &           \\
			&           &           &           & -t^\prime &   -\mu    & -t^\prime \\
			&           &           &           &           & -t^\prime &   -\mu
		\end{bmatrix},\\
	S=&\begin{bmatrix}
		0   &  \Delta &        &         &         &         &        \\
		-\Delta &    0    & \Delta &         &         &         &        \\
		& -\Delta & 0      &  \Delta &         &         &        \\
		&         & \ddots &  \ddots &  \ddots &         &        \\
		&         &        & -\Delta &    0    &  \Delta &        \\
		&         &        &         & -\Delta &    0    & \Delta \\
		&         &        &         &         & -\Delta &  0
	\end{bmatrix},
\label{Eq31}
\end{split}
\end{equation}
with $t^\prime = t + i \gamma$. Energy spectrum of the system can be obtained by diagonalizing the Hamiltonian $H_{1}^{(0)}$. As shown in Fig.~\ref{Fig4}(a), in the absence of the dissipative couplings, we find that the energy spectrum of the Kitaev model is divided into two different regimes. When parameter $\mu$ satisfies the condition $0\le\mu<2t$, the system is in the topologically non-trivial phase with two MBS. When $\mu\ge2t$, the system is in topologically trivial phase, which means that zero modes no longer exist. However, if the dissipative coupling term $\hat{U}_1$ is taken into account, the conclusions for the Kitaev model are changed.

Figure~\ref{Fig4} shows the variations of real parts for eigenenergies of the system when parameter $\mu$ is varied. Here, we assume that the Kitaev chain is composed of 40 qubits. The red solid curves in Figs.~\ref{Fig4}(a) and (c) are enlarged in Figs.~\ref{Fig4}(b) and (d), respectively. We find that dissipative couplings significantly change the topological properties of system. If the dissipative couplings do not exist, as we see from Figs.~\ref{Fig4}(a) and (b), a pair of zero-energy modes exists in topologically non-trivial regime and the corresponding energy splitting increases monotonously. However, if considering the dissipative couplings, as shown in Figs.~\ref{Fig4}(c) and (d), the topological phase transition point shifts towards the left of the value $\mu=2t$, which is the phase transition point when dissipative couplings do not exist. That is, the dissipative couplings make topologically non-trivial regime get narrower. Moreover, the energy splitting of zero-energy modes becomes oscillatory with the variation of $\mu$ when there exist dissipative couplings. If the dissipative coupling strength is further increased, then the phase transition point of the system shifts further to the left of the point $\mu=2t$, as shown in Figs.~\ref{Fig4}(e) and (f). Here, the phase transition point in Fig.~\ref{Fig4}(f) is not well displayed due to the finite length of the chain. In the thermodynamic limit, i.e., $N\rightarrow\infty$,  the topological phase transition point is $\mu = 2t/\sqrt{1+\gamma^2/t^2}$. This result can be obtained via the Hamiltonian in the momentum space, which will be described in detail later.

\begin{figure}
	\includegraphics[scale=0.2]{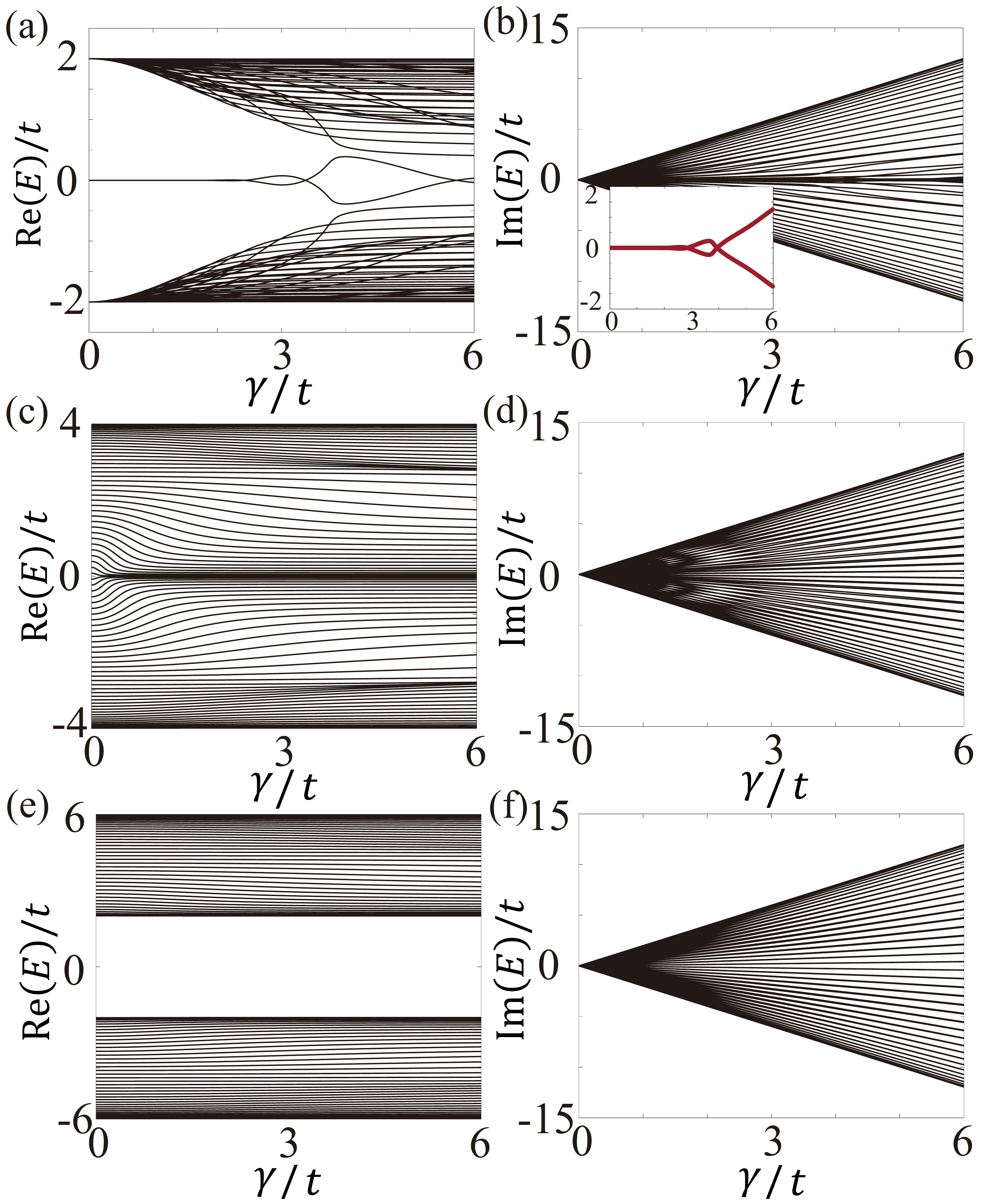}
	\caption{Energy spectrum of the Kitaev model $\hat{H}_1$ as a function of $\gamma$ for three different parameter regimes, under open boundary condition. (a) and (b) represent topologically non-trivial regime with $\mu=0$; (c) and (d) represent topological phase transition point with $\mu/t = 2$; (e) and (f) represent topologically trivial regime with $\mu/t = 4$. The inset in (b) shows enlarged imaginary parts of eigenvalues for edge states. Other parameters are the same as those in Fig.~\ref{Fig2}.}\label{Fig5}
\end{figure}

For the conventional Kitaev model Hamiltonian $\hat{H}_{0}$ in Eq.~(\ref{Eq23}),  the topologically non-trivial regime is $\mu<2t$, the trivial regime is $\mu>2t$ and the topological phase transition point is $\mu=2t$. To study the effect of  dissipative couplings on topological properties of the system, the real and imaginary parts of eigenvalues are plotted as a function of $\gamma$ in these three regimes. As shown in Fig.~\ref{Fig5},  we choose parameter $\mu=0$, $4t$ and $2t$, which correspond to topologically non-trivial, trivial regimes and topological phase transition point, respectively. We find that the dissipative couplings have different effects on the real part of the energy spectrum for these three parameter regimes. As shown in Fig.~\ref{Fig5}(a), it is obvious that real parts which correspond to the energies of zero-energy modes oscillate as a function of the parameter $\gamma$ in topologically non-trivial regime. Moreover, with the increasing of dissipative coupling strength $\gamma$, the amplitude of oscillation becomes larger. As shown in Fig.~\ref{Fig5}(b), the behavior of the  imaginary parts corresponding to the zero-energy modes is similar to those of the real parts. That is, eigenvalues of MBS are complex and exhibit oscillatory real and  imaginary parts. Moreover, the dissipative couplings also make difference to the localization of the system. Energy splitting of MBS becomes larger with the increase of $\gamma$. This means that the localization of these states get weaker due to the dissipative couplings. However, the effect of the dissipative couplings on the properties of the Kitaev chains in topologically trivial phase and topological phase transition point is very weak. When $\mu\ge2t$, as shown in Figs.~\ref{Fig5}(c) and (e), we find that the real parts of energy spectrum are barely changed. For imaginary parts of the energy spectrum, as shown in Figs.~\ref{Fig5}(b), (d) and (f), each state corresponds to a unique imaginary part,  there are no obvious differences for the imaginary parts in these three parameter regimes.

\begin{figure}
	\includegraphics[scale=0.22]{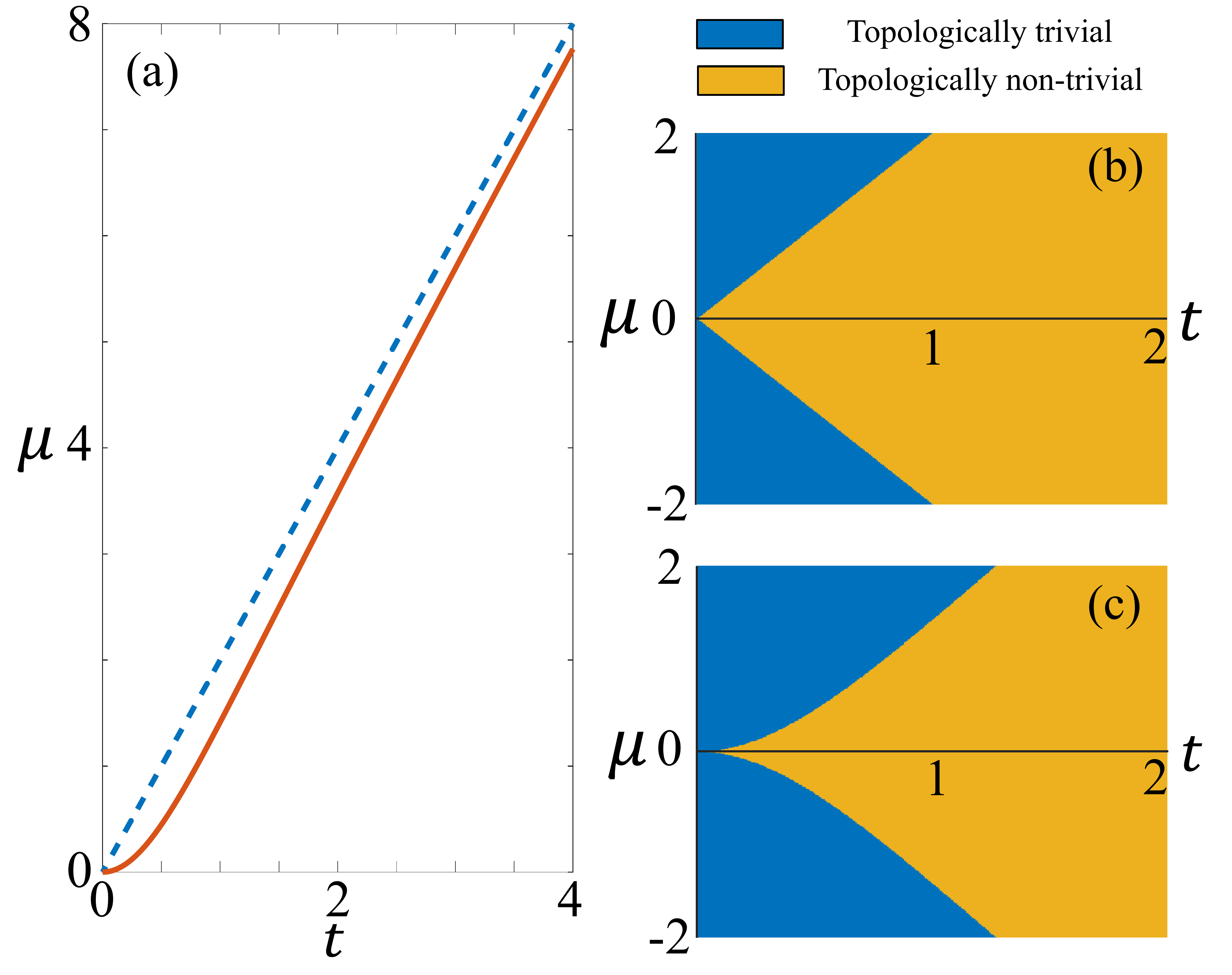}
	\caption{(a) Topological phase transition point for $\hat{H}_{1}$ without (blue dashed curve) and with (orange solid curve) dissipative couplings. (b) and (c) represent the phase diagrams for $\hat{H}_{1}$ without and with dissipative couplings. Dissipative coupling strength in (a) and (c) is fixed at $\gamma/t = 1$.}\label{Fig6}
\end{figure}

From Fig.~\ref{Fig4}, it is clearly shown that the topological phase transition point is changed by dissipative couplings. Taking similar procedure shown in Sec.~\ref{subsub4a2}, the analytical expression of the phase transition point with dissipative coupling can be obtained. Under the periodic boundary condition, $\hat{H}_1$ in Eq.~(\ref{Eq29}) can be written as
\begin{equation}
	\hat{H}_{1} =\frac{1}{2}\sum_{k} \hat{\Phi}_k^\dagger H_{k,1} \hat{\Phi}_k,
	\label{Eq32}
\end{equation}
in the momentum space with
\begin{equation}
H_{k,1} = 2\Delta \sin k\,\sigma_y - [\mu + 2(t + i\gamma)\cos k] \sigma_z.
	\label{Eq33}
\end{equation}
By setting eigenvalues of $H_{k,1}$ to be zero, we can obtain the phase transition condition
\begin{equation}
	\mu = \dfrac{2t}{\sqrt{1+\gamma^2/t^2}} 
	\label{Eq34}
\end{equation}
when $\Delta = t$.  It is clear that this condition can be reduced to the conventional case $\mu=2t$ when there is no dissipative coupling, which is shown as the blue dashed curve in Fig.~\ref{Fig6}(a). After introducing the dissipative couplings $\gamma$, the topological phase transition condition is determined by Eq.~(\ref{Eq34}), in which $\mu$ and $t$ have a nonlinear relation, and this relation is shown as the orange solid curve in Fig.~\ref{Fig6}(a). From the phase diagrams shown in Figs.~\ref{Fig6}(b) and (c), we can find that dissipative couplings induced by the common environment compress the topologically non-trivial region.

\subsubsection{Topology of the Kitaev model with the dissipative couplings for  $\theta \neq 0$ \label{subsub4b2}}

We now study the topological properties of the Kitaev model with the dissipative couplings for all nearest neighboring qubits when $\theta \neq 0$, which is schematically shown in Fig.~\ref{Fig3}(b). In this case, the Hamiltonian of the system is $\hat{H}_{\rm nonH}$, which is given in Eq.~(\ref{Eq22}).
We find that the couplings between neighboring qubits in Eq.~(\ref{Eq22}) are nonreciprocal when $\theta \neq 0\ \rm {or}\  \pi$. And this nonreciprocal coupling  results in new properties of the system. Below, we mainly study the edge and bulk properties from the energy spectrum and localization.

\begin{figure}
\includegraphics[scale=0.22]{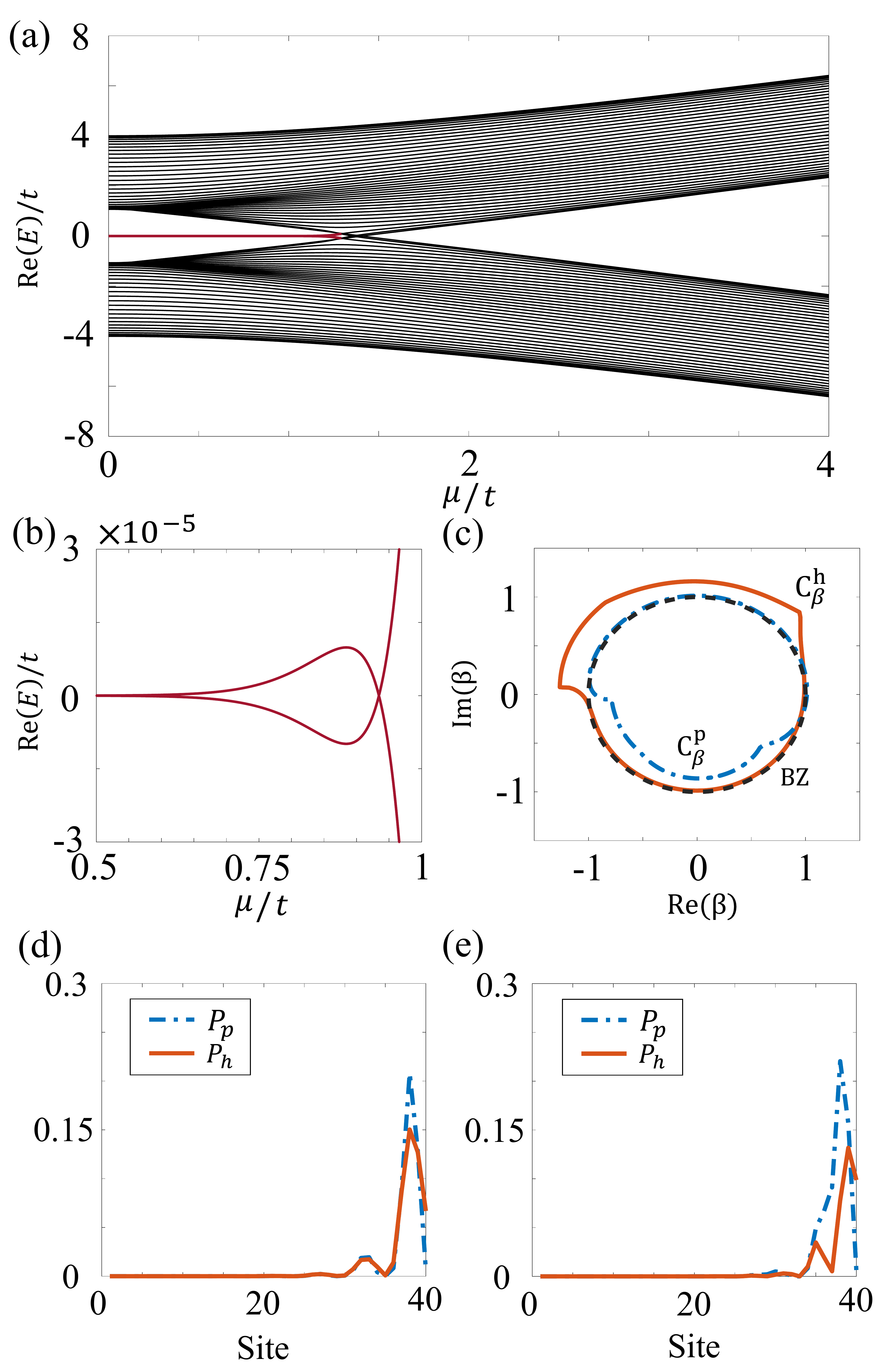}
\caption{(a) The real part of the energy spectrum for the Hamiltonian $\hat{H}_{\rm nonH}$ as a function of the parameter $\mu$, the red solid curves are enlarged in (b). (c) The corresponding GBZ for (a) when $\mu/t=0.2$. (d) and (e) represent the population distributions of the bulk states with the energy ${\rm Re}(E)/t=1.13$ and ${\rm Re}(E)/t=1.07$ when $\mu/t=0.2$, respectively. The blue dash-dotted and orange solid curves correspond to the particles and holes, respectively. Other parameters chosen here are $\Delta/t=1.3$, $\gamma/t=0.2$, $\theta=\pi/4$, $N=40$, and $t=1$.}\label{Fig7}
\end{figure}

For edge states, as shown in Fig.~\ref{Fig7}(a), dissipative couplings play the same role to the system as $\theta = 0$. That is, the dissipative couplings result in the oscillation of the energy corresponding to edge-states, which is clearly shown in Fig.~\ref{Fig7}(b), and the topological phase transition point also shifts towards the left of the point $\mu=2t$.

As for the bulk states, we use generalized Brillouin zone (GBZ)~\cite{PhysRevLett.121.086803,PhysRevLett.123.246801} to study the bulk-boundary correspondence of the system. Here, GBZ is the extension of Brillouin zone. Under Hermitian case, it is a unit circle, which means that the population of bulk states distributes over the whole system. However, for non-Hermitian case, GBZ may not be a unit circle. This means that the system becomes localized and the population of the bulk states may concentrate at the edges of the system and exhibit the skin effect.

For general situation, i.e., $\Delta \neq t$, we can rewrite the Hamiltonian$\hat{H}_{\rm nonH}$ in the momentum space as
\begin{equation}
\hat{H}_{\rm nonH} =\frac{1}{2}\sum_{k} \hat{\Phi}_k^\dagger H_{k,2}(e^{ik}) \hat{\Phi}_k,
\label{Eq35}
\end{equation}
 with the Bloch Hamiltonian
\begin{equation}
	\begin{split}
	H_{k,2}(e^{ik}) = h_{I,2}(e^{ik})I +h_{y,2}(e^{ik}) \sigma_y +
	 h_{z,2}(e^{ik})\sigma_z,
	\end{split}
	\label{Eq36}
\end{equation}
in which
\begin{eqnarray}
	h_{I,2}(e^{ik}) &=& it\sin\theta (e^{ik}-e^{-ik}), \nonumber \\
	h_{y,2}(e^{ik}) &=& - i \Delta(e^{ik}-e^{-ik}),  \\
	h_{z,2}(e^{ik}) &=& - \mu - t\cos\theta(e^{ik}+e^{-ik}) -  i\gamma(e^{ik}+e^{-ik}). \nonumber	\label{Eq37}
\end{eqnarray}
Extending the real wave number $k$ to complex plane, e.g., replacing $e^{ik}$ by $\beta$, we can regard the non-Bloch Hamiltonian $H_{k,2}(\beta)$  as a generalization of the Bloch Hamiltonian. Then the Hamiltonian in Eq.~(\ref{Eq36}) becomes
\begin{equation}
	H_{k,2}(\beta) = h_{I,2}(\beta)I + h_{y,2}(\beta) \sigma_y +
	h_{z,2}(\beta)\sigma_z,
	\label{Eq38}
\end{equation}
which has eigenvalues
\begin{equation}
	E_{\pm} = h_{I,2}(\beta) \pm \sqrt{h_{y,2}^2(\beta) + h_{z,2}^2(\beta)}.
	\label{Eq39}
\end{equation}
Here, the parameters in Eq.~(\ref{Eq39})  are given as
\begin{equation}
	\begin{split}
	h_{I,2}(\beta) &= it\sin\theta (\beta-\beta^{-1}), \\
	h_{y,2}(\beta) &= - i \Delta(\beta-\beta^{-1}), \\
	h_{z,2}(\beta) &= -\mu - t\cos\theta(\beta+\beta^{-1}) - i\gamma(\beta+\beta^{-1}).
	\end{split}
\label{Eq40}
\end{equation}
The GBZ can be obtained via the characteristic equation
\begin{equation}
	f(\beta,E_\pm) = \det[H_{k,2}(\beta)-E_\pm] = 0.
	\label{Eq41}
\end{equation}
This equation has four solutions $|\beta_1| \le |\beta_2| \le |\beta_3| \le |\beta_4|$. Choosing the solution of $\beta$ which manifests $|\beta_2| = |\beta_3|$, the trajectories of $\beta_2$ and $\beta_3$ construct GBZ. Figure~\ref{Fig7}(c) shows the GBZ of the nonreciprocal Kitaev model $\hat{H}_{\rm nonH}$, where the two loops correspond to particles and holes, respectively, and they satisfy $\beta_h$ = $\beta_p^{-1}$. It is clear that these two loops are not unit circles. This means that the population of particles and holes mainly concentrate at the edge of the system, i.e., skin effect, and this conclusion can be confirmed from Figs.~\ref{Fig7}(d) and (e), which show the population distribution of partial bulk states. Under open boundary condition, these states demonstrate that particles and holes are all localized at the boundary. Similarly, other bulk states also become localized and mainly concentrate towards the edge of the system.

\subsubsection{Topology of the Kitaev model with the dissipative couplings for special positions of the chain  when $\theta = 0$ \label{subsub4b3}}

\begin{figure}
	\includegraphics[scale=0.2]{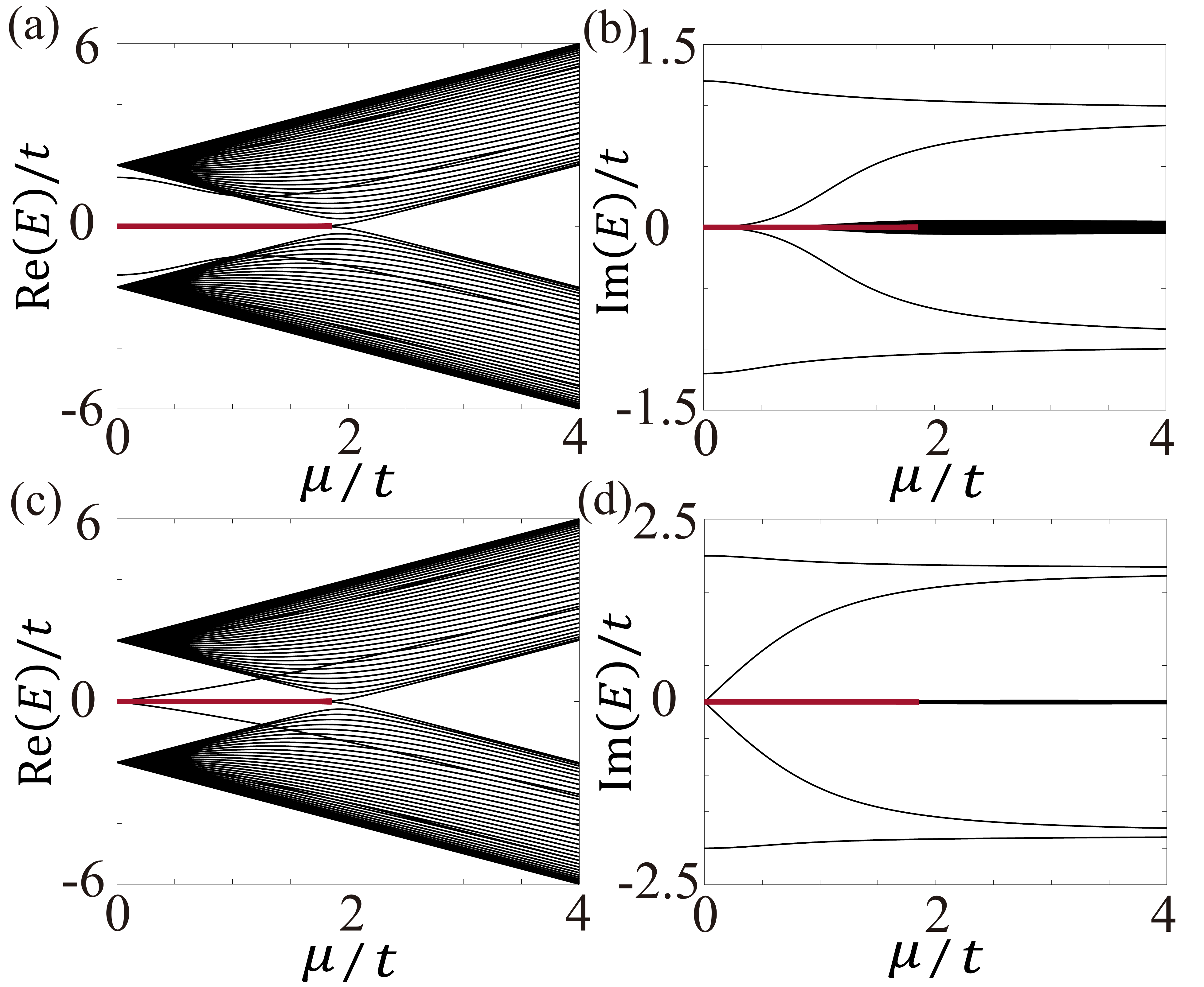}
	\caption{Energy spectrum of the Hamiltonian $\hat{H}_2(1)$ as a function of the parameter $\mu$ for different $\gamma$, under open boundary condition. (a) and (b) $\gamma/t = 1.2$, (c) and (d)  $\gamma/t = 2$. The red solid curves correspond to energies of edge states. Other parameters are the same as those in Fig.~\ref{Fig2}.}\label{Fig8}
\end{figure}

In this section, we discuss the cases that the dissipative couplings only exist in two pairs of the qubits, as schematically shown in Figs.~\ref{Fig3}(c) and (d).  We first assume that the dissipative couplings only exist in qubit pairs at the two ends of the Kitaev chain, then the effective Hamiltonian can be derived from Eq.~(\ref{Eq29}) as
\begin{equation}
	\begin{split}
		\hat{H}_2(1) = \hat{H}_{0} - i\gamma \left( \hat{a}_{1}^\dagger \hat{a}_{2} + \hat{a}_{N}^\dagger \hat{a}_{N-1} + {\rm H.c.} \right).
	\end{split}
\label{Eq42}
\end{equation}
The energy spectra as a function of the parameter $\mu$ for different dissipative coupling strengths are shown in Fig.~\ref{Fig8}. Hereafter, in this subsection, we take $N=40$.

For $\mu<2t$, in the case of relatively weak dissipative coupling, as shown in Fig.~\ref{Fig8}(a), we find that the real parts of energy spectrum are similar to those of the Hermitian case shown in Fig.~\ref{Fig4}(a). There is a pair of zero-energy states that can be regarded as boundary states, and their energies are real despite the presence of the non-Hermitian dissipative couplings. However, from Fig.~\ref{Fig8}(b), we can find the occurrence of imaginary parts for the energies corresponding to these bulk states, and this shows that some bulk states are changed by such dissipative coupling. Further increasing the dissipative coupling strength, we can see from Figs.~\ref{Fig8}(c) and (d) that there exist four additional modes whose energies are around zero in topologically non-trivial phase. These modes possess complex eigenvalues, which means that these states are unstable  in the topologically non-trivial phase.

For  $\mu>2t$, as shown in Fig.~\ref{Fig8}, the dissipative couplings almost make no effect on the real parts of the energy spectrum. Also, the imaginary parts of energy spectrum for most states are nearly zero although there exist dissipative couplings. 

\begin{figure}
\includegraphics[scale=0.2]{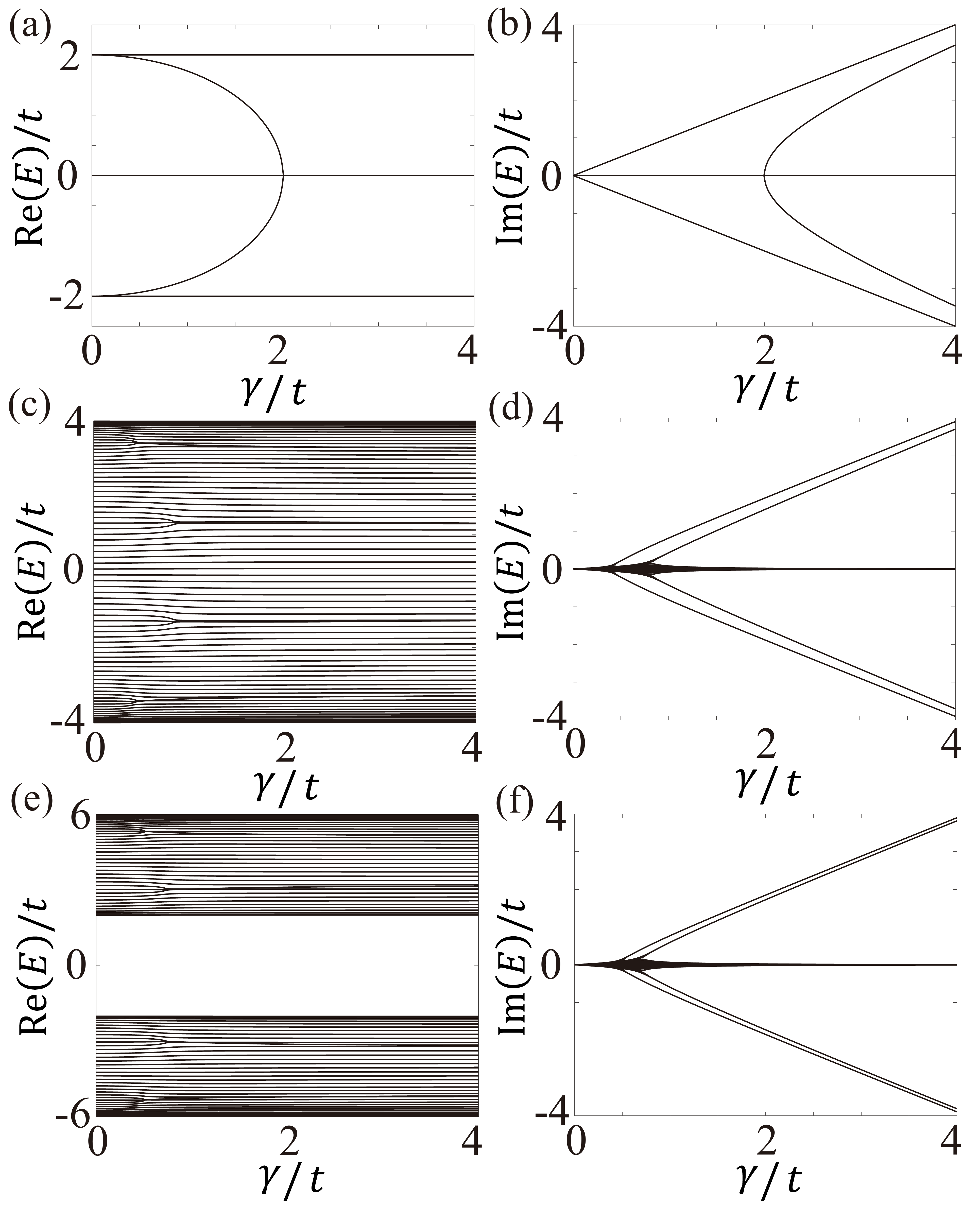}
\caption{Energy spectrum of the Kitaev model $\hat{H}_2(1)$ as a function of $\gamma$ for different $\mu$, under open boundary condition. (a) and (b) $\mu/t = 0$, (c) and (d) $\mu/t = 2$, (e) and (f)  $\mu/t = 4$. Other parameters are the same as those in Fig.~\ref{Fig2}.}\label{Fig9}
\end{figure}

It is clear that the dissipative couplings have different effects on the energy spectrum in different parameter regimes. As shown in Figs.~\ref{Fig9}(a) and (b), for $\mu=0$, there exist four states, whose energies are always complex values when $\gamma \neq 0$, the real parts of these energies are equal to $2t$. Moreover, when $0\le\gamma<2t$, there exist four states whose energies are purely real and monotonically close to zero. When $\gamma\ge2t$, the energies corresponding to these four bulk states become purely imaginary. That is, energies of these four bulk states are either real or imaginary depending on the strength of dissipative coupling $\gamma$. For edge states, we find that energies of MBS in topologically non-trivial phase are always purely real. This means that the edge states have strong robustness against the common environment which locate at both ends of the system. For $\mu=2t$ and $\mu=4t$, as shown in Figs.~\ref{Fig9}(c)-(f), we find that dissipative coupling barely changes the real parts of the energy spectrum. However, eigenvalues of almost all states have imaginary parts, most of which are around zero. Thus we conclude that the dissipative couplings at two ends change the energy spectrum of the system, especially for topologically non-trivial regime.

The dissipative coupling can also exist in other two pairs of the nearest neighboring qubits, e.g., the pair between the $m$th and $(m+1)$th qubit, and the pair between the $(N-m)$th and $(N-m+1)$th qubits. Here, we consider the case that $\theta=0$. In this case, the corresponding effective Hamiltonian can be written as
\begin{equation}
	\begin{split}
	\hat{H}_2(m) = &\ \,\hat{H}_{0} + \hat{U}_2(m),\\
	\hat{U}_2(m) = &-i\gamma \left( \hat{a}_{m}^\dagger \hat{a}_{m+1} + \hat{a}_{N-m}^\dagger \hat{a}_{N-m+1} + {\rm H.c.} \right).
	\label{Eq43}
	\end{split}
\end{equation}
Here, for convenience, we call $m$ as the dissipative coupling position.

\begin{figure}
	\includegraphics[scale=0.21]{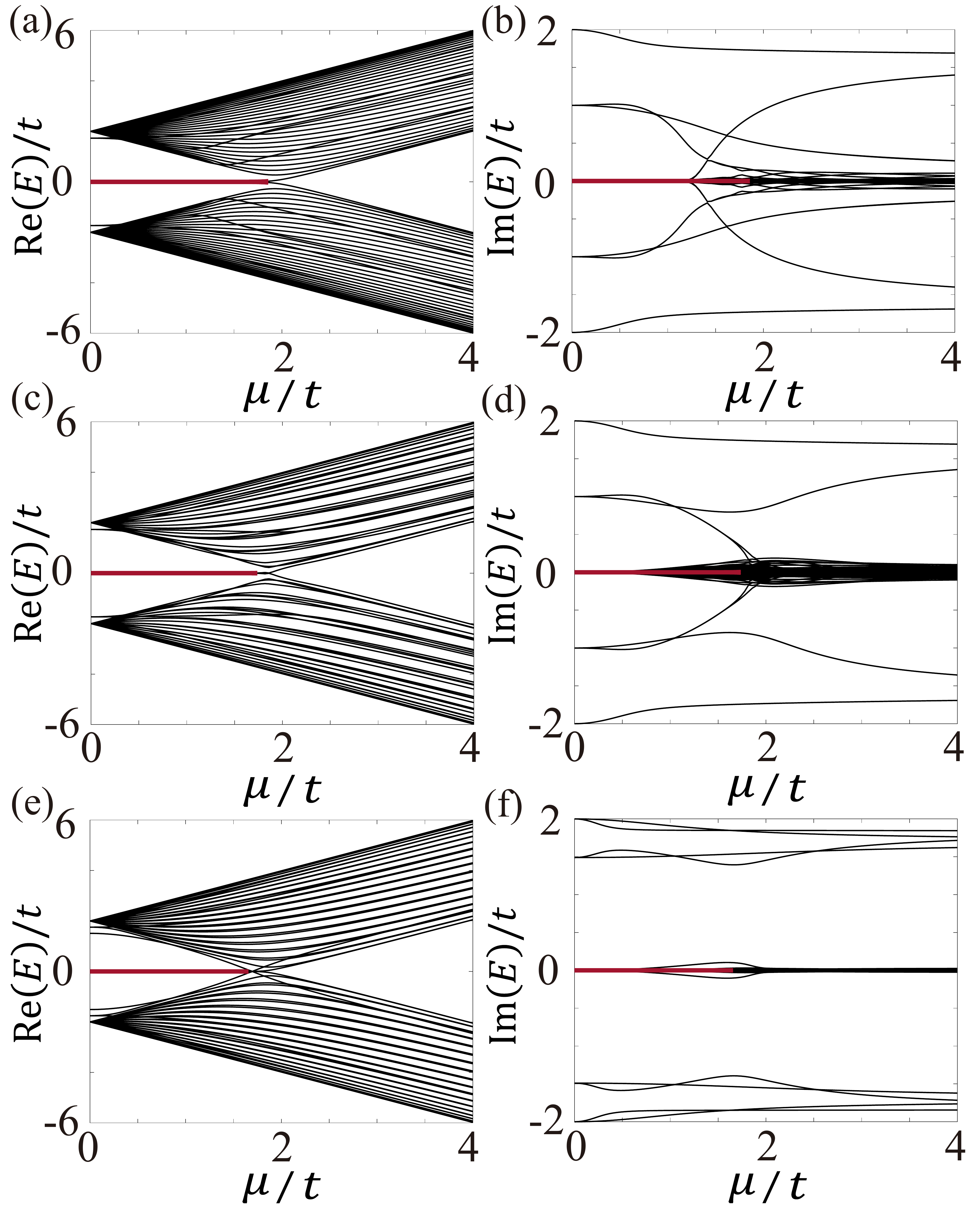}
	\caption{Energy spectrum of the Kitaev model $\hat{H}_2(m)$ as a function of $\mu$ for different coupling position $m$, under open boundary condition.  (a) and (b) $m = 2$, (c) and (d) $m = 10$, (e) and (f) $m = 19$. The red solid curves correspond to energies of edge states. Other parameters are taken as $\gamma/t=2$, $\Delta/t=1$,  $N=40$, and $t=1$.}\label{Fig10}
\end{figure}

Below, we study the effect of the dissipative coupling position $m$ on the topological properties of the system. As shown in Figs.~\ref{Fig10}(a)-(f), the coupling position mainly changes the energy spectrum structure in topologically non-trivial phase, especially for bulk states. For real parts of energy spectrum, dissipative couplings mainly change the energies of the lowest (highest) energy state in the upper (lower) band of the bulk states. Fig.~\ref{Fig8}(c) corresponds to the case $m=1$, besides edge states, there exist at least four modes whose real parts of eigenenergies are near zero. With the increase of $m$, the real parts of eigenenergies corresponding to these four modes are away from zero, which are shown in Figs.~\ref{Fig10} (a), (c) and (e). When the dissipative couplings distribute around the center of the qubit chain, more eigenenergies have nonzero imaginary parts, but the energies corresponding to edge states are always purely real. Therefore, we can conclude that the edge states are robust against the dissipative couplings $\hat{U}_2(m)$.

\begin{figure}
	\includegraphics[scale=0.22]{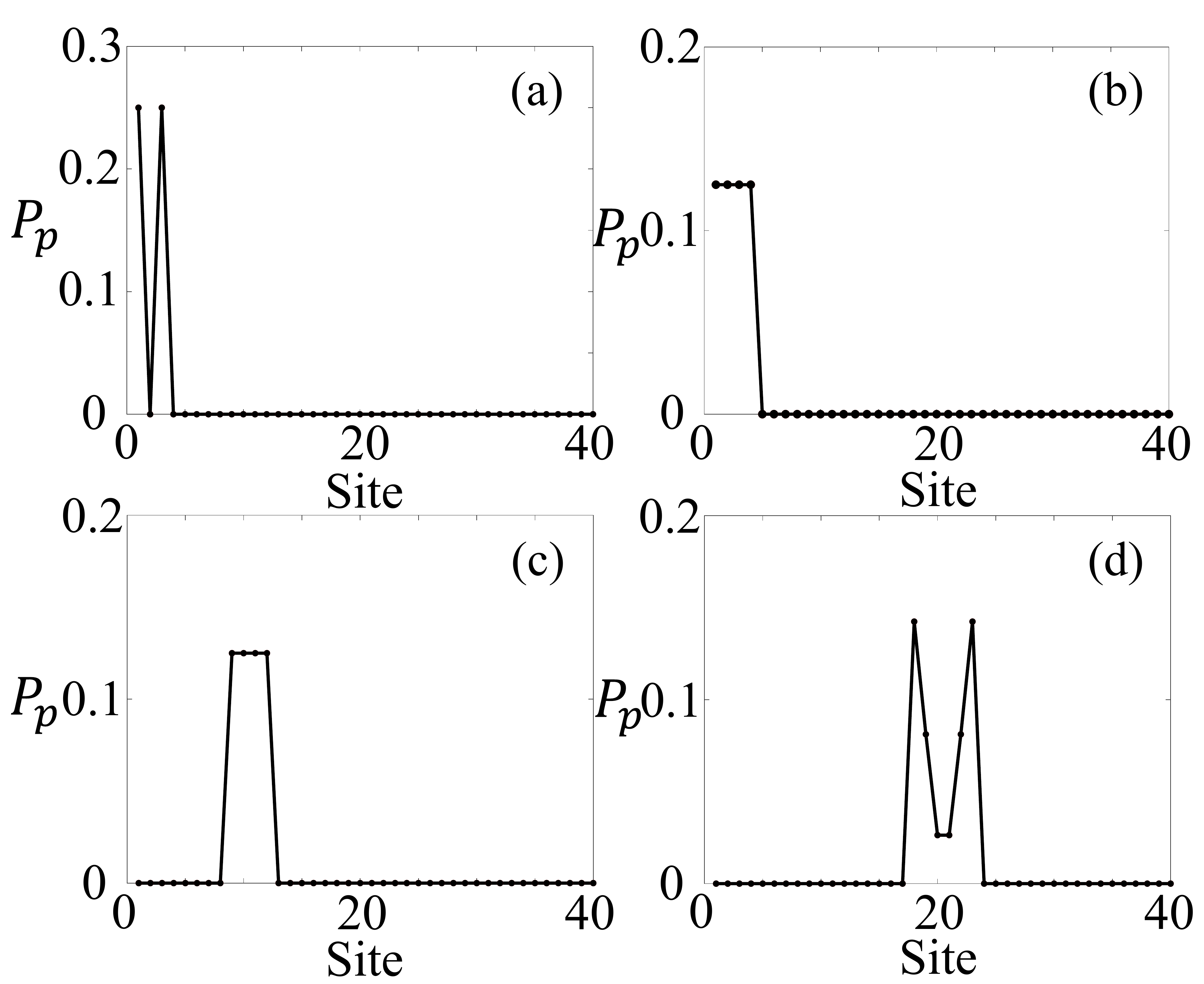}
	\caption{Population distribution of bulk states of $\hat{H}_2(m)$ when $\mu/t=0$ under open boundary condition. (a) $m = 1$, (b) $m = 2$, (c) $m = 10$, and (d) $m = 19$. Other parameters are the same as those in Fig.~\ref{Fig10}.}\label{Fig11}
\end{figure}

Moreover, we find that the localization of the system and the energy difference between edge states and the lowest (highest) energy state in the upper (lower) band of the bulk states are related to the dissipative coupling position. When $\mu=0$, the position $m=1$ corresponds to the smallest energy difference, and the lowest energy state in the upper band of the bulk states have the strongest localization, as shown in Fig.~\ref{Fig11}(a), the population distribution of lowest energy states in the upper band of the bulk states concentrate at the two sites of the left edge. When the dissipative coupling position is around the center of the qubit chain, the corresponding energy difference becomes wider and the localization of the lowest energy states in the upper band of the bulk states becomes weaker. Figs.~\ref{Fig11}(b) and (c) show the population distribution of the lowest energy states in the upper band of the bulk states for $m = 2$ and $m=10$, respectively. Four neighboring sites have the same population and distribute around the $2$th and $10$th sites. Due to the symmetry of the system, there exists a special dissipative coupling position, i.e., $m=19$. In this case, the localization of bulk states is strengthened compared with the other dissipative coupling positions except $m = 1$. As shown in Fig.~\ref{Fig11}(d), two peaks of the population distribution appear in the $18$th and $23$th sites.

\begin{figure}
	\includegraphics[scale=0.28]{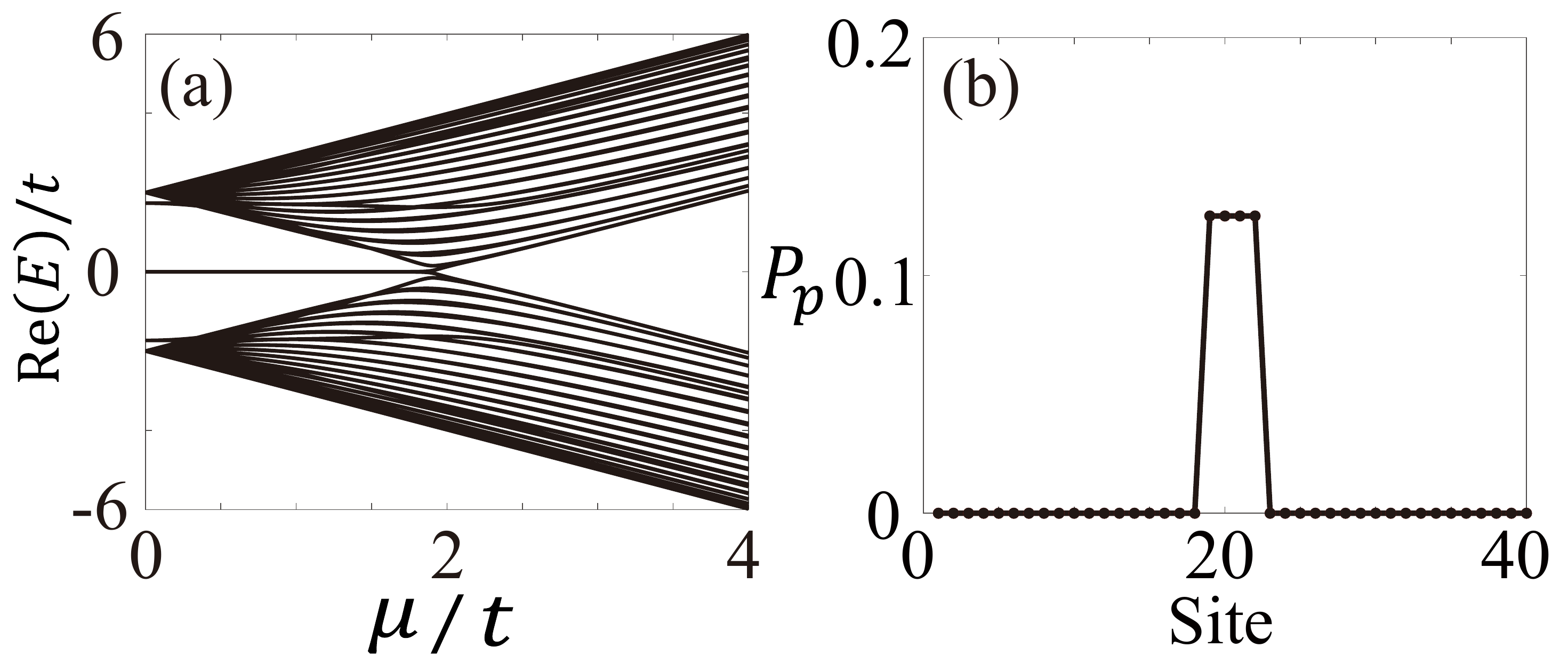}
	\caption{(a) The real part of energy spectrum of $\hat{H}_2(20)$ as a function of $\mu$. (b) The population distribution of bulk state of $H_2(20)$ when $\mu/t=0$. Other parameters are the same as Fig.~\ref{Fig10}.}\label{Fig12}
\end{figure}

Finally, we consider the case of $m = 20$. In this case, the dissipative coupling exists only between the $20$th and the $21$th qubits. We find that such dissipative coupling does not produce any special effects compared to the cases $m = 2$ and $10$. Energy spectrum structure in Fig.~\ref{Fig12}(a) is the same as those in Figs.~\ref{Fig10}(a) and (c). The corresponding population distribution of the lowest energy state in the upper band of the bulk states is shown in Fig.~\ref{Fig12}(b), which is also similar to those in Figs.~\ref{Fig11}(b) and (c). This state only populates the sites $19$, $20$, $21$, and $22$.

\subsubsection{Local environmental effect \label{subsub4b4}}

All we discussed above focus on the effect of dissipative coupling, but the onsite dissipative potential can also affect the topological property of system. The effect of onsite dissipative potential and comparison between dissipative coupling and onsite dissipative potential will be discussed below. In order to distinguish onsite dissipative potential and dissipative coupling, we use $\delta$ to represent this potential.

The onsite dissipative potential can also nontrivially change the topological phase of system~\cite{PhysRevA.94.022119}. If we only consider onsite dissipative potential, the Hamiltonian can be written as
\begin{equation}
	\begin{split}
		\hat{H}_3 &= \hat{H}_0 + \hat{U}_3, \\
		\hat{U}_3 &= -i\delta\sum_{n=1}^{N} \hat{a}_n^\dagger \hat{a}_n, \label{Eq44} 
	\end{split}
\end{equation}
with $\hat{H}_0$ given in Eq.~(\ref{Eq23}), the topological phase transition point is
\begin{equation}
	\mu = 2t\sqrt{1-\frac{\delta^2}{4\Delta^2}}. \label{Eq45}
\end{equation}
This equation shows that the onsite dissipative potential affects the topological phase transition. When $\delta<|2\Delta|$, there exists topologically non-trivial phase, along with edge states, while $\delta>|2\Delta|$, the topological non-trivial phase disappears no matter how parameter $\mu$ varies.
\begin{figure}
	\includegraphics[scale=0.2]{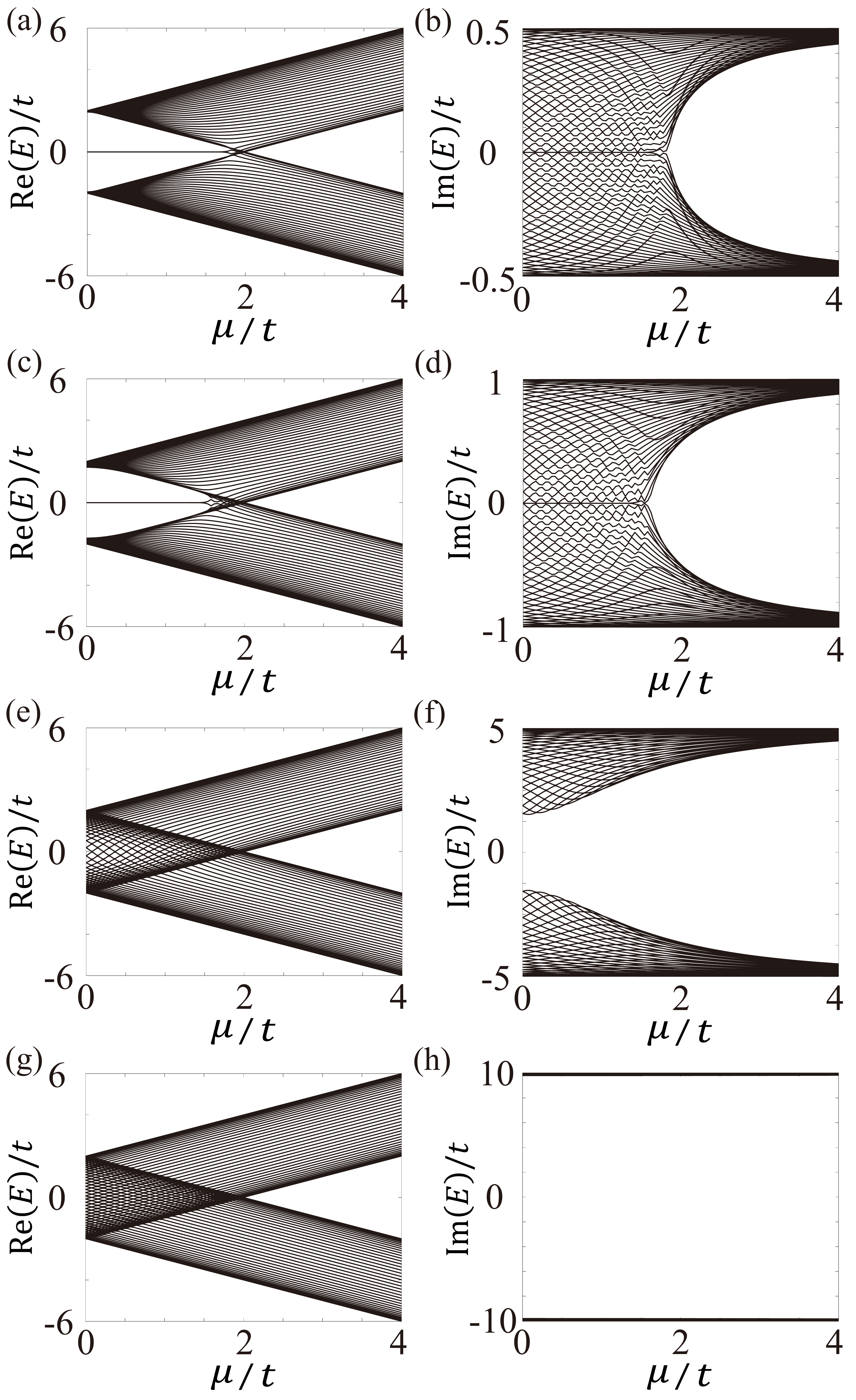}
	\caption{Energy spectrum of the Hamiltonian $\hat{H}_3$ as a function of the parameter $\mu$ for different $\delta$, under open boundary condition. (a) and (b) $\delta/t = 0.5$, (c) and (d) $\delta/t = 1$, (e) and (f) $\delta/t = 2$, (g) and (h) $\delta/t = 10$. Other parameters are the same as those in Fig.~\ref{Fig2}.}\label{Fig13}
\end{figure}

As shown in Figs.~\ref{Fig13}(a)-(d), when onsite dissipative potential $\delta$ is weak, e.g., $\delta/t = 0.5$ and $1$, we can find that there exist zero-energy states, the imaginary parts of energies for these states are zero, this means that the energies of these states are purely real when the system are stay in topologically non-trivial phase. However, for topologically trivial phase, energies of all states are complex. Further increase onsite dissipative  potential, e.g., $\delta/t = 2$ and $10$, in these cases, as we can see from Figs.~\ref{Fig13}(e)-(h), edge states no longer exist, along with the disappearance of topologically non-trivial regime. Besides that, energies of all states own imaginary parts, which means that energies of all states are complex.

Comparing Eq.~(\ref{Eq34}) with Eq.~(\ref{Eq45}), we can find that both onsite dissipative potential and dissipative coupling can nontrivially change topological phase, these two mechanisms make the topologically non-trivial regime get narrower. Besides that, energies of zero-energy states are purely real in these two cases. However, for nonzero dissipative coupling, there always exist topologically non-trivial regime no matter how large the parameter $\mu$ becomes when the onsite dissipative potential is zero. For nonzero onsite dissipative potential, this result will be modified when the dissipative coupling is zero, once $\delta>|2\Delta|$, the topologically non-trivial regime will disappear, which means that the system stays in the topologically trivial phase for arbitrary $\mu$.

Furthermore, we consider a more general model which takes both onsite dissipative potential and dissipative coupling into account at the same time. In this case, the Hamiltonian reads
\begin{equation}
	\hat{H}_4 = \hat{H}_0 + \hat{U}_1 +  \hat{U}_3. \label{Eq46}
\end{equation}
\begin{figure}
	\includegraphics[scale=0.2]{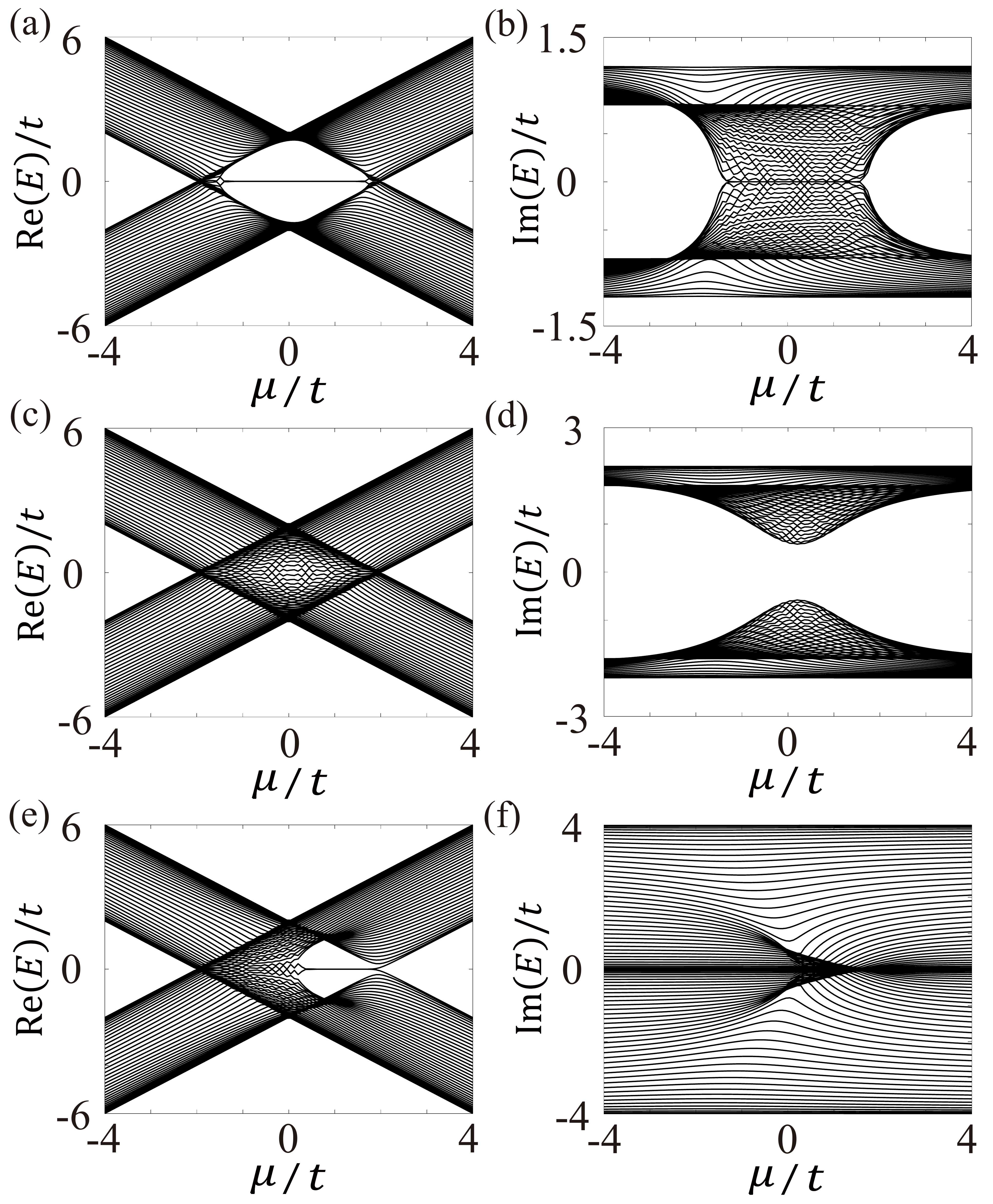}
	\caption{Energy spectrum of the Hamiltonian $\hat{H}_4$ as a function of the parameter $\mu$ for different $\delta$ and $\gamma$, under open boundary condition. (a) and (b) $\gamma/t = 0.1, \delta/t = 1$, (c) and (d) $\gamma/t = 0.1, \delta/t = 2$, (e) and (f) $\gamma/t = 1, \delta/t = 2$. Other parameters are the same as those in Fig.~\ref{Fig2}.}\label{Fig14}
\end{figure}
As shown in Fig.~\ref{Fig14}, energy spectrum of the system will be changed a lot if we combine onsite dissipative potential and dissipative coupling into the  Hamiltonian $\hat{H}_0$. First, the symmetry of energy spectrum no longer exists. Viewing from the real parts, for onsite dissipative potential, the energy gap occurs if the parameter $\mu$ manifests $-2t\sqrt{1-\delta^2/4\Delta^2} < \mu < 2t\sqrt{1-\delta^2/4\Delta^2}$, in this case, it is clear that the energy gap distributes around the origin symmetrically. For the dissipative coupling, the energy gap occurs if the parameter $\mu$ manifests $-2t/\sqrt{1+\gamma^2/t^2} < \mu < 2t/\sqrt{1+\gamma^2/t^2}$, in this case, it is clear that the energy gap also distributes around the origin symmetrically. Considering both onsite dissipative potential and dissipative coupling together, as shown in Figs.~\ref{Fig14}(a), (c) and (e), the interval where the energy gap exists becomes small and no longer distributes around the origin symmetrically, but the energy splitting oscillation exists in this case. For imaginary parts, as shown in Figs.~\ref{Fig14}(b), (d) and (f), the symmetry also beaks down, energy spectrum is not symmetric around the origin. Besides that, dissipative coupling expands the condition that topologically non-trivial phase can exist. As shown in Eq.~(\ref{Eq45}) which only includes the onsite dissipative potential, if $\delta \ge |2\Delta|$, the system will always be in topologically trivial regime for arbitrary $\mu$. However, the existence of dissipative coupling breaks this constraint, the topologically non-trivial regime can also occur even if $\delta \ge |2\Delta|$. As shown in Fig.~\ref{Fig13}(e), the non-trivial regime will not occur if the dissipative coupling does not exist, taking this coupling into account, as shown in Figs.~\ref{Fig14}(c) and (e), the non-trivial regime will come up again if the dissipative coupling is large enough.

\section{SUMMARY AND DISCUSSION \label{sec5}}

We theoretically propose and construct a Kitaev chain model based on superconducting qubits coupled through rf SQUID couplers via the flexible design of the superconducting quantum devices. Using two kinds of couplers between the neighboring qubits, we can realize a complex hopping parameter $te^{i\theta}$.  The pairing coupling parameter $\Delta$, which is  necessary for the Kitaev model and usually neglected as the counter-rotating terms in the study of the quantum optics, can be obtained via the frequency matching between time-dependent magnetic fluxes through the couplers and the coupled pairs of the qubits.  Also, the coupling strength between the qubits can be tuned by the magnetic fluxes through the loops of the rf SQUID couplers.  The Kitaev model, described by the fermiomic operators, is finally obtained by mapping the qubit operators to the fermiomic creation and annihilation operators via Jordan-Wigner transformations.

Actually, the environments cannot be avoided for all quantum systems. Thus, in our paper, we study the environmental effect. In particular, the nearest neighboring qubits or sites usually experience a common environment which can be mimicked by the microwave cavity or waveguide in superconducting quantum circuits and other solid-state quantum devices. Therefore, we mainly study the common environmental effect on the topological states of the Kitaev model. We find that the effect of common environment can be equivalent to the non-Hermitian dissipative couplings between nearest neighboring qubits with onsite dissipative potential. The effect of the onsite dissipative potential on the topological properties is also compared with the dissipative coupling.

By analyzing the energy spectrum and population distributions of the eigenstates corresponding to the effective Hamiltonian of the system, we find that the dissipative couplings have significant impacts on the topological properties of the Kitaev models. If the dissipative couplings exist for all neighboring sites, then they nontrivially change the population distribution of eigenstates and the topological phase transition point. If the dissipative couplings only exist in specific sites, then the number of near-zero energy states and the imaginary part of the energy spectrum vary with the variation of the dissipative coupling sites. We also find that the dissipative couplings at the edges have the significant impact on the lowest (highest) energy state in the upper (lower) band of the bulk states. Moreover, we  find that the Majorana bound states with zero energy exhibit strong robustness against the perturbations caused by common environment. We mention that the energy spectrum and population distribution can be measured in superconducting qubit circuits via, e.g., reflection spectrum of the measurement of cavities or waveguides.

Compared with other systems, superconducting qubit systems have mature fabrication process, good scalability and easy controllability for both the couplings and the energy structures of the qubits ~\cite{NatRevPhys.1.19, Rev.Mod.Phys.85.623,PhysRep.718.1}. Therefore, the superconducting qubit systems can serve as one of the most potential platforms for the research of topology.  For concreteness of our study, we use phase qubits as an example to construct the Kitaev model, however, the method can also be applied to other kinds of superconducting qubits. In our study, the coherence of the qubits is not very important, while the tunability of both qubit frequencies and  the couplings between qubits is crucial. We know that the tunability is not a difficult task for all kinds of superconducting qubit circuits. Our work may inspire researchers to apply this high-quality platform to conduct more researches in the field of topology. We also hope that our studies for the effect of common environment on the topology can further stimulate people to do more researches on the relation between the environment and topology.

\section{Acknowledgments}

Y.X.L. is supported by NSFC  under Grant No. 11874037 and the Key R \&D Program of Guangdong province under Grant No. 2018B030326001.

\appendix

\end{document}